\documentclass[article]{achemso}
\usepackage[utf8]{inputenc}

\usepackage{graphicx}
\usepackage{graphicx}
\usepackage{bm}
\usepackage{amsmath}
\usepackage{amssymb}
\usepackage{appendix}
\usepackage{comment}
\usepackage{relsize}
\usepackage[bottom]{footmisc}
\usepackage[dvipsnames]{xcolor}
\usepackage[section]{placeins}
\usepackage{layouts}
\usepackage{bbold}
\usepackage{enumitem}
\usepackage{subcaption}
\usepackage{braket}
\usepackage{threeparttable}
\usepackage{tabularx}
\usepackage{multirow}

\usepackage{soul} 

\definecolor{darkblue}{rgb}{0,0,.65}
\definecolor{darkgreen}{rgb}{0.28,0.41,0.19}

\usepackage{braket}
\usepackage[version=3]{mhchem}
\usepackage{tikz}
\usepackage{hyperref}
\usepackage{url}
\newcommand{\imag}{\mathrm{i}}

\title{Energy-filtered excited states and real-time dynamics served in a contour integral}
\author{Ke Liao}
\email{ke.liao.whu@gmail.com}
\affiliation{Faculty of Physics, Arnold Sommerfeld Centre for Theoretical
Physics (ASC),\\Ludwig-Maximilians-Universit{\"a}t M{\"u}nchen,
Theresienstr.~37, 80333 M{\"u}nchen, Germany} 
\alsoaffiliation{Munich Center for
Quantum Science and Technology (MCQST), Schellingstrasse 4, 80799 M{\"u}nchen,
Germany}
\alsoaffiliation{Max Planck Institute for Solid State Research, Heisenbergstraße 1, 70569 Stuttgart}

\begin{document}

\begin{abstract}
    It is observed that the Cauchy integral formula (CIF) can be used to represent
    holomorphic functions of diagonalizable operators on a finite domain. This
    forms the theoretical foundation for applying various operators in the form
    of a contour integral to a state, while filtering away eigen-components that
    are not included by the contour. As a special case, the identity operator in
    the integral form--the Riesz projector--is used to design an
    algorithm for finding a given number of eigen-pairs whose energies are close
    to a specified value in the equation-of-motion coupled cluster singles and
    doubles (EOM-CCSD) framework, with applications to calculate core excited
    states of molecules which is relevant for the X-ray absorption spectroscopy
    (XAS). As a generalization, I showcase a novel real-time electron dynamics
    (RT-EOM-CCSD) algorithm based on the CIF form of the exponential time-evolution operator, 
    which admits extremely large time steps while preserving
    accurate spectral information.
\end{abstract}

\maketitle

\section{Introduction}
While significant progress has been made in studying excited states of molecules 
using quantum chemical methods over the past several decades, certain challenges persist, 
particularly in developing methods to efficiently find eigenstates near a specified energy 
value and in performing real-time dynamics that can selectively target specific portions of the eigen-spectrum.
Early response theories based on real-time 
dynamics in Hartree-Fock (HF) and density functional theory (DFT)~\cite{mclachlan1964,schonhammer1978,runge1984,akama2010} 
established fundamental approaches to calculating excited states of molecules. 
Subsequently, similar response theories have been developed in the coupled cluster (CC) 
framework~\cite{monkhorst1977,dalgaard1983,koch1990,stanton1993,bartlett2007b,krylov2008,
bartlett2012,sonk2011,kvaal2012,lopata2012,pigg2012a,helgaker2014,
nascimento2016,nascimento2017,pedersen2019,kristiansen2020,skeidsvoll2020,sverdrupofstad2023}.
In the meantime, specialized techniques have also emerged to target specific excited states more directly. 
These include the 
core-valence separation approaches~\cite{cederbaum1980,coriani2015,vidal2019} for core excitations, 
energy-specific scheme~\cite{peng2015}, 
and various shift-and-invert techniques~\cite{yu2017,dorando2007a,booth2012,zhao2019}. 
Complementary to these methods, frequency-domain approaches based on the correction vector 
method~\cite{kuhner1999,mcclain2016b,lewis2019a,lee2023b}, 
damped response~\cite{kristensen2009,schnack-petersen2023} and 
complex polarization propagator approaches~\cite{norman2001,norman2005,ekstrom2006,coriani2012,coriani2012a,pedersen2014} 
also allow direct calculation of spectral features at selected frequencies. 

Based on the contour integral representation of the identity 
operator—the Riesz projector~\cite{riesz2012}, 
the FEAST algorithm first introduced in the context of 
density functional theory (DFT)~\cite{polizzi2009} provides 
a systematic approach to find eigen-states within a specified energy window. 
However, a prior estimate of the number of eigen-states in the window is necessary 
for its efficiency and stability~\cite{dinapoli2016}, which can be challenging when
the energy window is densely populated. Despite its success in DFT and as a general 
eigen-value solver, its adoption in quantum chemistry has been relatively limited, 
with notable applications primarily in density matrix renormalization group (DMRG)~\cite{baiardi2022b} 
for calculating vibrational states.

In this work,  I propose an algorithm for finding $n$ eigen-states with 
energies near a specified value in the framework of equation-of-motion coupled cluster 
singles and doubles (EOM-CCSD) theory with an adaption of the FEAST algorithm.
This is achieved by dynamically adjusting the contour 
radius to include the desired number of eigen-states among a fixed number of trial vectors. 
This strategy differs from the original FEAST algorithm~\cite{polizzi2009} or its 
DMRG extension~\cite{baiardi2022b}, which aims to find all eigen-states within a 
fixed energy interval. To demonstrate the algorithm's potential, 
I apply it to study core-excited states relevant to X-ray absorption spectroscopy
 (XAS), which are crucial for understanding molecular electronic structure.

More importantly, I demonstrate that the Riesz projector used in FEAST 
is a special case of the Cauchy integral formula (CIF), which can represent
general holomorphic functions of diagonalizable operators. This observation
enables much broader potential theoretical and algorithmic developments. I notice that 
quantum algorithms based on the CIF have been
proposed in Ref.~\cite{takahira2020}, however, only with a focus on the quantum
resource estimation and error analysis. In this work,
I provide a concrete application by using the CIF to represent the time-evolution operator, 
resulting in a novel real-time electron dynamics algorithm in the 
EOM-CCSD framework that allows for significantly larger time steps while
preserving accurate spectral information. This algorithm effectively filters
out unwanted eigen-components from higher and lower energy states irrelevant
to the dynamics of interest, making it valuable for evolving states with dominant
components in densely clustered energy windows (like valence or core excited regions).
The CIF approach opens numerous potential applications across quantum chemistry, 
condensed matter physics, and quantum information science, 
e.g.~thermal expectation value calculations within 
specified energy windows~\cite{hummel2018,white2018,harsha2019}, 
calculating long-time averages of observables within narrow energy
windows—central to verifying the eigenvalue thermalization hypothesis
(ETH)~\cite{deutsch1991,srednicki1994,rigol2008,deutsch2018}, 
and state preparation for quantum simulations~\cite{irmejs2024}.

The paper is organized as follows. In the Theory section, I
introduce the theoretical background of the CIF and provide a brief
recapitulation of the EOM-CCSD method. I then present its FEAST extension,
which enables the calculation of excited states around any energy value.
Additionally, I explain how the CIF-based real-time EOM-CCSD can be
implemented. The details of the computational setup and the technical tools used
in this work are outlined in Computational Details. In the
 Results and Discussion section, I demonstrate the capability of
the FEAST-EOM-CCSD algorithm by calculating the core excited states of several
molecules and present a preliminary study of the CIF-based real-time electron
dynamics of the H$_2$O molecule. Finally, I conclude with a summary and discuss
potential future directions for CIF-based algorithms.

\section{Theory}
\label{theory}
The Cauchy integral formula (CIF) is a powerful tool in complex analysis, and it
states that for any holomorphic function $f(z)$ on the complex plane, the value
of $f(a)$ at a point $a$ inside a simple closed curve $C$ can be calculated by
the following formula
\begin{equation}
    f(a) = \frac{1}{2\pi \imag}\oint_{C}\frac{f(z)}{z-a}\mathrm d{z}. \label{equ:cauchy}
\end{equation}
A function $f(z)$ is called holomorphic when it is differentiable at every point
in its domain. 
When $f(z)=1$, the operator version of Eq.~\eqref{equ:cauchy} defines the Riesz
projector~\cite{riesz2012}
\begin{equation}
    \hat{P}\equiv \sum_{p\in\mathcal{C}}\ket{\Psi_p}\bra{\Psi_p} = \frac{1}{2\pi \imag}\oint_{C}\frac{1}{z\hat{I}-\hat{A}}\mathrm d{z}, \label{equ:projector}
\end{equation}
where $\hat{I}$ is the identity operator and $\hat{A}$ is in general a hermitian
matrix or a hermitian operator,~e.g. the Hamiltonian, with $\ket{\Psi_p}$ being
its eigen-vectors. I use the letter $\mathcal{C}$ to indicate the set of
indices of eigen-values circled by the contour $C$. This crucial identity
enables the so-called FEAST algorithm to calculate the eigen-values and
eigen-states within a specified energy window in the context of density
functional theory~\cite{polizzi2009} and in matrix product states related
algorithms~\cite{baiardi2022b}. I observe that the above identity
Eq.~\eqref{equ:projector} is a special case and that CIF can be used to
represent general holomorphic functions of diagonalizable operators on a finite
domain as 
\begin{equation}
    f(\hat{A}) = \frac{1}{2\pi \imag}\oint_{C}\frac{f(z)}{z\hat{I}-\hat{A}}\mathrm d{z},
    \label{equ:cif_A}
\end{equation}
where $\hat{A}$ is a diagonalizable operator and can be written in its spectral
decomposition as $\hat{A}=\sum_ia_i\ket{\Psi^{R}_{i}}\bra{\Psi^{L}_i}$ using its
right- and left-eigen-vectors. It can easily be shown that for all $z$ on the
contour $C$ such that $|z|>||\hat{A}||$, where $||\hat{A}||$ can be any matrix
norm, the Taylor expansion of the resolvent $1/(z\hat{I}-\hat{A})$ is
well-defined. This corresponds to the condition that the contour $C$ encloses
all the eigen-values of $\hat{A}$. Therefore, using the eigen-decomposition of
$\hat{A}$ and the normal CIF, one can arrive at the above formula
Eq.~\eqref{equ:cif_A}. Alternatively, if one is only interested in a part of the
eigen-spectra of $\hat{A}$, one can define a contour $C$ on the complex plane 
that encloses only the
eigen-values of interest. This is equivalent to defining a new matrix
$\tilde{A}=\sum_{p\in \mathcal{C}}a_p\ket{\Psi^{R}_{p}}\bra{\Psi^{L}_p}$
containing parts of the original eigen-components selected by the contour $C$
and indicated by $p\in{\mathcal C}$.

Based on these theoretical insights, I will show that (i) the original FEAST
algorithm based on the projector Eq.~\eqref{equ:projector} can be adapted to the
EOM-CCSD theory involving a non-Hermitian Hamiltonian 
 with small modifications; and (ii) a novel real-time electron
dynamics algorithm in the EOM-CCSD framework can be developed based on the
integral form of the time-evolution operator $e^{-\imag\bar{H}t}=\frac{1}{2\pi
\imag}\oint_{C}\frac{e^{-z}}{z\hat{I}-{\imag\bar{H}t}}\mathrm d{z}$. 

\subsection{EOM-CCSD}
In this work, I focus on the neutral electronic excitations of a system, and I
recapitulate accordingly the electronic excitation version of EOM-CCSD
(EE-EOM-CCSD) theory in this subsection. I note that the theories discussed in
the following subsections can be easily adapted to other versions of EOM-CCSD
and potentially other methods as well, such as DFT, other quantum chemical
methods and even some tensor network theories. In the rest of the paper, I will
refer to the EE-EOM-CCSD theory as EOM-CCSD for simplicity.

The EOM-CCSD theory solves the following effective eigen-value problem
\begin{equation}
    \bar{H}\ket{\Psi_k} = E_k\ket{\Psi_k},
    \label{equ:eom}
\end{equation}
where $E_k$ is the  $k$th excitation energy
 corresponding to the $k$th excited state $\Psi_k$. The
non-Hermitian effective Hamiltonian $\bar{H}$ is defined as
\begin{equation}
    \bar{H} = e^{-\hat{T}}\hat{H}e^{\hat{T}}-E_{\rm CCSD},
\end{equation}
where $E_{\rm CCSD}$ is the CCSD ground state energy and 
$\hat{T}=\sum_{ai}T^a_i\hat{a}^{\dagger}_a\hat{a}_i+1/4\sum_{abij}T^{ab}_{ij}\hat{a}^{\dagger}_a\hat{a}^{\dagger}_b\hat{a}_j\hat{a}_i$
is the cluster operator containing the singles and doubles operators, whose
amplitudes are determined by solving the CCSD amplitudes equations, and
$\hat{H}$ is the second quantized Hamiltonian in a given set of orbital basis
$\{\phi_s\}$, defined as
\begin{equation}
    \hat{H} = \sum_{pq}h_{pq}\hat{a}^{\dagger}_p\hat{a}_q + \frac{1}{4}\sum_{pqrs}V_{pqrs}\hat{a}^{\dagger}_p\hat{a}^{\dagger}_q\hat{a}_s\hat{a}_r.
\end{equation}
In the above, I follow the convention that the indices $p,q,r,s$ run over all
 the orbitals, $a,b,c,d$ run over the virtual orbitals, and $i,j,k,l$ run over
 the occupied orbitals. The EOM-CCSD eigen-vectors $\ket{\Psi_k}$ 
can be written as
\begin{equation}
    \ket{\Psi_k} = \ket{D_0}+ \sum_{ai}u^a_i \hat{a}^{\dagger}_a\hat{a}_i \ket{D_0} + \frac{1}{4}\sum_{abij}u^{ab}_{ij}\hat{a}^{\dagger}_a\hat{a}^{\dagger}_b\hat{a}_j\hat{a}_i \ket{D_0},
\end{equation}
where $\ket{D_0}$ is the reference determinant, normally the Hartree-Fock
determinant. The $u^a_i$ and $u^{ab}_{ij}$ are the amplitudes to be determined
by projecting the EOM-CCSD amplitudes equation Eq.~\eqref{equ:eom} to the
singles and doubles subspace. Normally, the modified Davidson algorithm is used
to solve the non-Hermitian eigen-value problem Eq.~\eqref{equ:eom} for a few
low-lying eigen-states. I refer to
Refs.~\cite{emrich1981,emrich1981a,stanton1993,bartlett2007b,krylov2008,bartlett2012,helgaker2014} for more details on
the EOM-CCSD theory. In the following subsection, I will take advantage of one
of the key steps in the EOM-CCSD algorithm, namely the matrix-vector
multiplication $\bar{H}\ket{\Psi_k}$, to achieve FEAST-EOM-CCSD.

\subsection{FEAST-EOM-CCSD}
The key idea of the FEAST algorithm is to use the projector
Eq.~\eqref{equ:projector} to project trial vectors into a subspace spanned by
eigen-states whose energies are circled in by the contour $C$. Similarly,  for the
non-Hermitian $\bar{H}$ in EOM-CCSD, the biorthogonal left- and
right-eigen-states can be used to decompose the resolvent of $\bar{H}$.
Therefore, one can define the following projector 
\begin{equation}
\bar{P}= \sum_{p\in \mathcal{C}} \ket{\Psi^R_p}\bra{\Psi^L_p} = \frac{1}{2\pi \imag}\oint_{C}\frac{1}{z\hat{I}-\bar{H}}\mathrm d{z},
\end{equation}
where $\bra{\Psi^L_p}$ and $\ket{\Psi^R_p}$ are left- and right-eigen-state
associated with the eigen-value $E_p$, and they fullfil
$\braket{\Psi^L_p|\Psi^R_q}=\delta_{pq}$. I point out that the biorthogonal
formulation is used  only for showing that the projector $\bar{P}$ can 
indeed filter eigen-components as in a Hermitian case. Since currently only
the excitation energies are of interest, I use only the right trial vectors to generate the
subspace for the non-Hermitian eigen-value problem, and as in normal one-sided EOM-CCSD
theory, no left eigen-vectors are calculated.

\begin{figure}[ht!]
    \centering
    \begin{subfigure}[b]{0.48\textwidth}
        \centering
        \includegraphics[width=\textwidth]{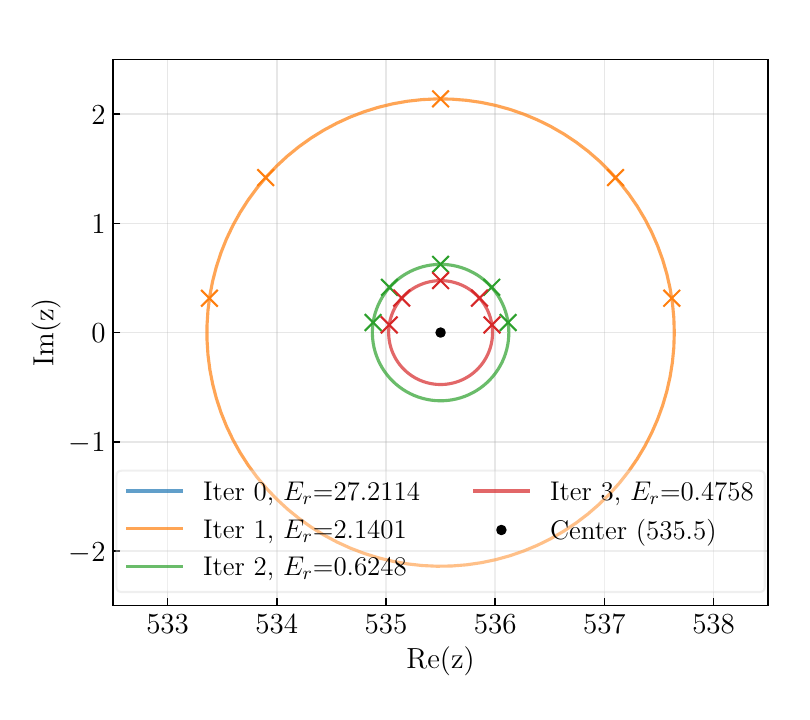}
    \end{subfigure}
    \hfill
    \begin{subfigure}[b]{0.48\textwidth}
        \centering
        \includegraphics[width=\textwidth]{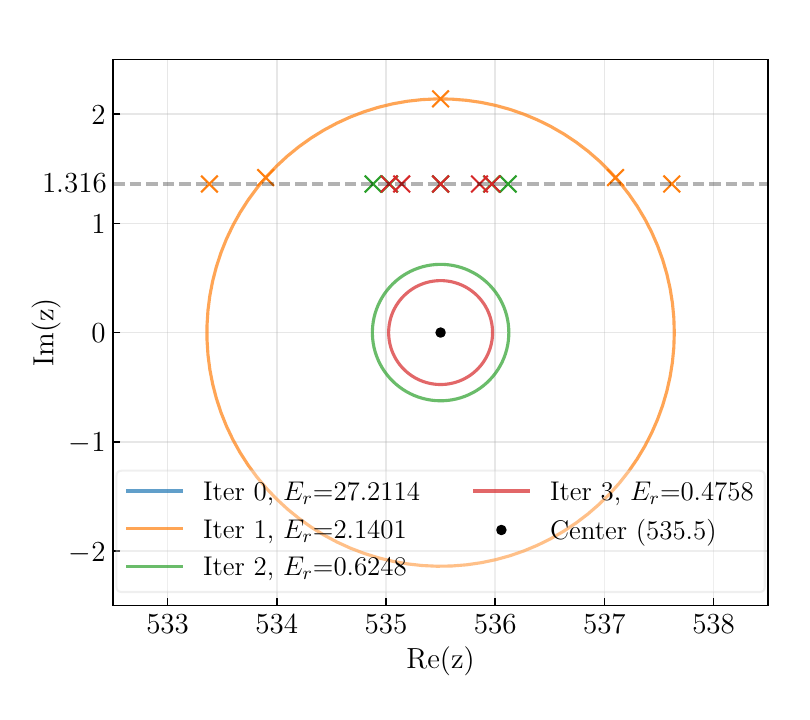}
    \end{subfigure}
    \begin{subfigure}[b]{0.48\textwidth}
        \centering
        \includegraphics[width=\textwidth]{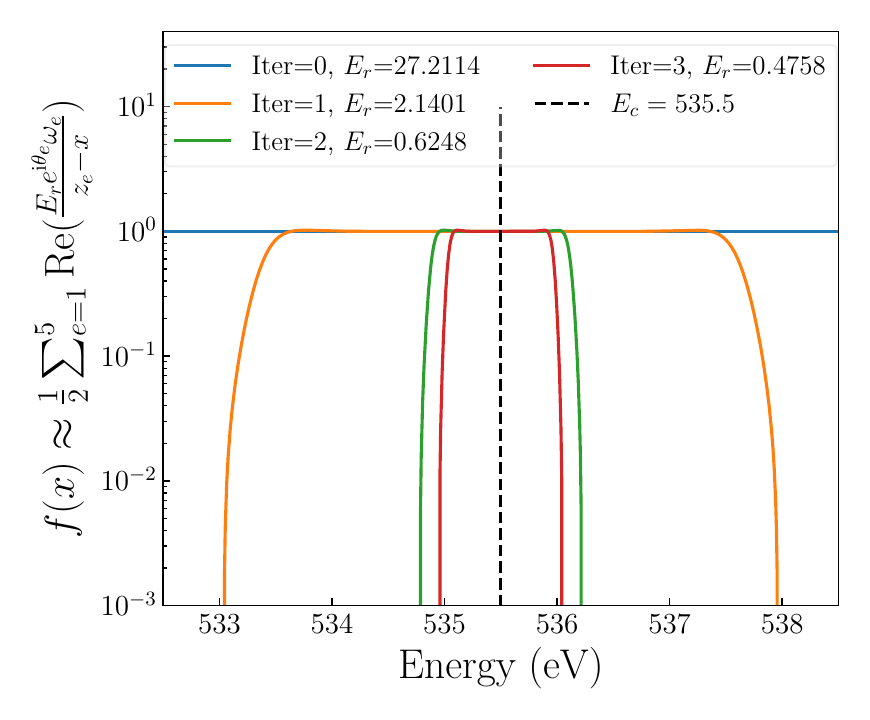}
    \end{subfigure}
    \hfill
    \begin{subfigure}[b]{0.48\textwidth}
        \centering
        \includegraphics[width=\textwidth]{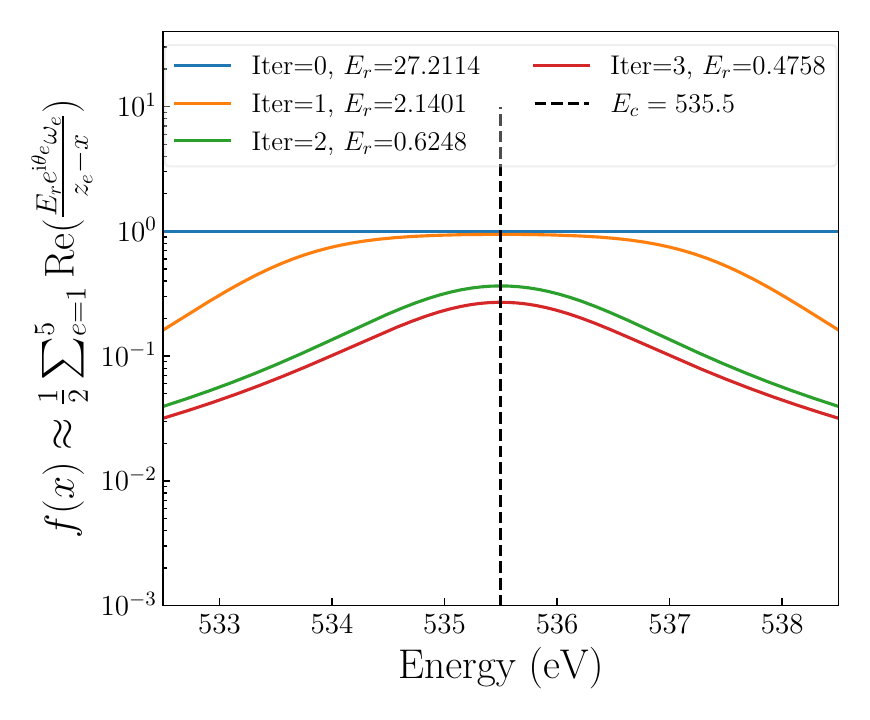}
    \end{subfigure}
    \caption{Energy filtering effect of applying the projector $\bar{P}$, demonstrated using 
    10 Gauss-Legendre quadrature nodes on the contour $C$ (only 5 nodes are shown and the conjugation relation is applied)
     with varying radii $E_{\rm r}$ while keeping 
    a constant number of eigen-states $n=5$ around the energy center $E_{\rm c}=535.5$ eV as 
    the iterations proceed.
    (Left) The filtering function $f(x)\approx \frac{1}{2}\sum_{e=1}^5\mathrm{Re}(\frac{E_r e^{\mathrm{i}\theta_e}\omega_e}{z_e-x})$. (Right)
    The same filtering function, but limiting the magnitude of the imaginary part of $z_e$ to be not smaller than 0.05 Ha (1.361 eV).}
    \label{fig:filter}
\end{figure}

Starting from a random trial vector $\ket{\Phi^{\rm T}}$, I can apply the
projector $\bar{P}$ on it to obtain the new state
\begin{equation}
    \ket{\tilde{\Phi}}=\sum_{p\in \mathcal{C}} \ket{\Psi^R_p}\braket{\Psi^L_p|\Phi^{\rm T}}=
     \frac{1}{2\pi \imag}\oint_{C}\frac{1}{z\hat{I}-\bar{H}}\ket{\Phi^{\rm T}}\mathrm d{z},
\end{equation}
which has support on $\{\ket{\Psi^R_p} |  p\in\mathcal{C}\}$ only. The two key
 steps are to apply $\frac{1}{z\hat{I}-\bar{H}}$ on $\ket{\Phi^{\rm T}}$ and to
 carry out the contour integral in an efficient and accurate way. They can be
 achieved by recasting the integral into a weighted summation along a set of $K$
 Gauss-Legendre quadrature nodes on the circle $C$, see the upper panel of Fig.~\ref{fig:filter},
\begin{equation}
    \ket{\tilde{\Phi}} \approx \frac{1}{2}\sum_{e=1}^{K/2}\omega_e \mathrm{Re}\left(E_r\mathrm{e}^{\mathrm i\theta_e}\ket{Q_e}\right),
\end{equation}
where $z_e$, $\theta_e$ and $\omega_e$ are the coordinate, the angle and the weight at quadrature node $e$, 
$E_r$ is the current radius of the contour,
and $\ket{Q_e}$ is the solution of the following linear system
\begin{equation}
    (z_e\hat{I}-\bar{H})\ket{Q_e} = \ket{\Phi^{\rm T}}.
\end{equation}

One can generate $n$ different trial vectors following the above recipe and
orthogonalize them~\footnote{In contrast to the original FEAST algorithm, which
does not require orthogonalization of the trial vectors, I find that using
orthogonalized trial vectors improves stability of the overall algorithm.}, with
which the matrix elements of ${\bf \bar{H}}^{\rm eff}$ in the subspace can be
calculated as
\begin{equation}
    \bar{H}^{\rm eff}_{ij} = \braket{\tilde{\Phi}_i|\bar{H}|\tilde{\Phi}_j}-E_{\rm CCSD}\delta_{ij}.
\end{equation}
Solving the following non-Hermitian eigen-value problem 
\begin{equation}
    {\bf\bar{H}^{\rm eff}}\ket{\Psi_p} = E_p\ket{\Psi_p}
    \label{equ:eff_eig}
\end{equation}
yields eigen-values $E_p$ within the desired energy window defined by the
contour $C$.

\begin{figure}[ht!]
    \centering
    \begin{subfigure}[b]{0.48\textwidth}
        \centering
    \includegraphics[width=\linewidth]{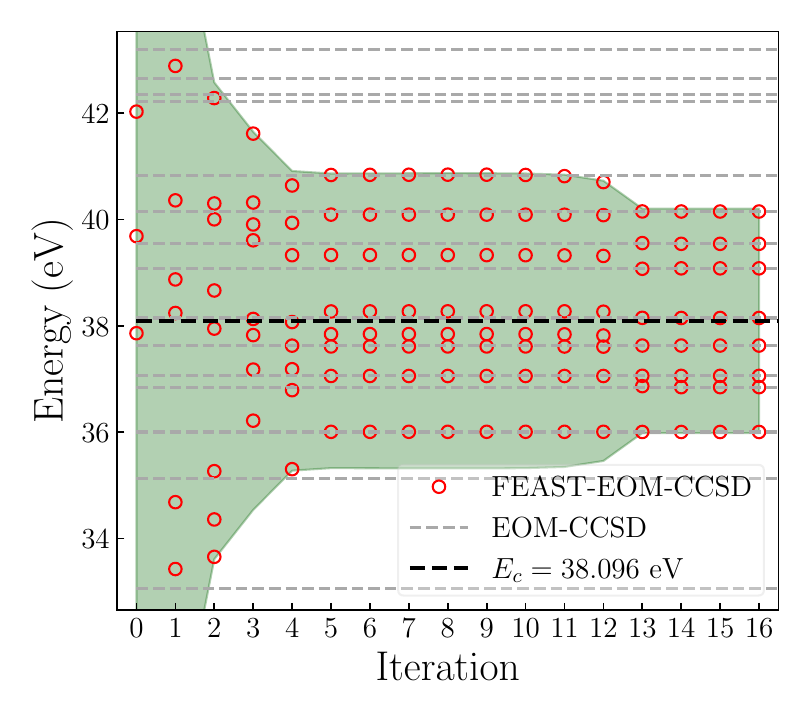}
    \end{subfigure}
    \begin{subfigure}[b]{0.48\textwidth}
        \centering
        \includegraphics[width=\textwidth]{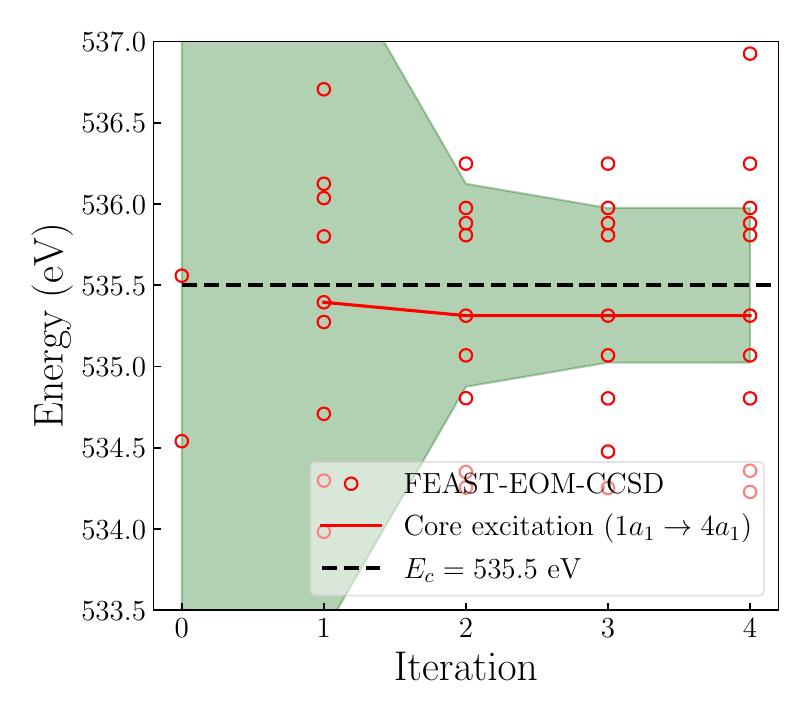}
    \end{subfigure}
    \caption{ FEAST-EOM-CCSD targetting low-lying and core
    excitation energies of the H$_{\rm 2}$O molecule. ( Left) Benchmarking FEAST-EOM-CCSD 
    against EOM-CCSD by calculating 8 low-lying excitation energies near 
    $E_c=1.4$ Ha using the STO-6G basis set. 
    (Right) FEAST-EOM-CCSD targetting 5 highly-excited states 
    around $E_c=535.5$ eV using the aug-cc-pVTZ basis set. The core excitation ($1a_1\to 4a_1$) energy is
    marked by the red line.
    The grey dashed lines are the reference
    EOM-CCSD eigen-values on the left figure and the red dots are the FEAST-EOM-CCSD
    eigen-values in both figures. The green shaded area is the dynamically adjusted energy window in both figures.} 
    \label{fig:h2o_feast}
\end{figure}

In contrast to the original FEAST algorithm~\cite{polizzi2009,dinapoli2016},
which relies on a prior estimate of the number of eigen-states within a
specified energy window for convergence and stability, I adopt a different
strategy of finding a given number of eigen-states whose energies are close to a
specified value. This is achieved by dynamically adjusting the contour radius
using the current estimate of eigen-values and the specified energy value. 
I outline the FEAST-EOM-CCSD algorithm in Table~\ref{tab:feast_algorithm}.
\begin{table*}[ht]
    \centering
    \caption{Steps of the FEAST-EOM-CCSD algorithm.}
    \begin{tabularx}{\textwidth}{lX}
        \hline
        Step & Description \\
        \hline
        1 & Choose an energy center $E_{\rm c}$ and the number $n$ of
        eigen-states to target around it. \\
        2 & Generate $n$ random trial vectors and optionally also a small number
         $m$ of supplemental random trial vectors, $\{\ket{\Phi^{\rm T}_i}\},
         i=1,\dots,n+m$, and choose the initial contour radius $E_{\rm r}$ to be
         a large value, e.g.~1 Ha. \\
        3 & Generate 5 (effectively K=10 as conjugation relation is used) Gauss-Legendre nodes $\{x_e\}$ between [-1,1] and
        their associated weights $\{w_e\}$; Calculate the complex nodes'
        coordinates by $z_e = E_{\rm c} + E_{\rm r} \exp(\imag \theta_e)$, where
        $\theta_e = -\frac{\pi}{2}(x_e - 1)$. \\
        4 & For each trial vector $\ket{\Phi^{\rm T}_i}, i=1,\dots,n+m$, solve
        the linear system problem $(z_e\hat{I}-\bar{H})\ket{Q_e} =
        \ket{\Phi_i^{\rm T}}$ at each quadrature node $e$ and obtain the new
        trial vector $\ket{\tilde{\Phi}_i} \leftarrow \frac{1}{2}\sum_{e=1}^{K/2}\omega_e 
        \mathrm{Re}\left(E_r\mathrm{e}^{\mathrm i\theta_e}\ket{Q_e}\right)$. \\
        5 & Use the QR decomposition~\cite{Harris2020} to orthogonalize the
        trial vectors $\{\tilde{\Phi}_i\}$ and calculate the ${\bf \bar{H}}^{\rm eff}$
        elements in the subspace spanned by $\{\ket{\tilde{\Phi}_i}\}$. \\
        6 & Diagonalize the ${\bf \bar{H}}$ of size $(n+m)\times(n+n)$ to obtain the
        eigen-values $\{\lambda_i\}$ and eigen-vectors $\{{\bf c}_i\}$,
        $i=1,2,\cdots,n+m$. \\
        7 & Update the trial vectors using the eigen-vectors: $\ket{\Phi^{\rm
        T}_i} \leftarrow \sum_{j=1}^nc_{i}^{j}\ket{\tilde{\Phi}_j}$. \\
        8 & Update $E_{\rm r}$ such that $n$ current eigen-values
        closest to $E_{\rm c}$ are circled, by calculating and sorting the distances $d=|\lambda_i-E_c|$
        and choosing the $n$th as the new $E_r\leftarrow d_n$. \\
        9 & Check if convergence is reached, i.e.~the change in the norm of the
        $n$ eigen-values between two consecutive cycles is smaller than a
        threshold $\epsilon$. If not, go to step 3. \\
        \hline
    \end{tabularx}
    \label{tab:feast_algorithm}
\end{table*}
In practice, I find that the more accurately the linear systems are solved, the
faster the algorithm converges. Typically, I use 5-20 iterations per linear
system, depending on how well-conditioned the linear systems are. 
The $m$ supplemental trial vectors are used to ensure that possible
degeneracies around the largest and smallest eigen-values are resolved for the
$n$ eigen-states. They also improve the overall convergence of the algorithm. 
  I  demonstrate the algorithm's effectiveness using the H$_2$O molecule with a minimal basis
    set in Fig.~\ref{fig:h2o_feast} (Left), targeting low-lying eigenvalues
    around a specific value of $E_c=38.096$ eV and comparing the results with reference
    EOM-CCSD eigenvalues. This benchmarks FEAST-EOM-CCSD's ability 
    to accurately target eigenvalues without calculating the lower-energy states. 
    Additionally, I show the convergence of 
    a core-excitation energy near $E_c=535.50$ eV for the H$_2$O molecule using the aug-cc-pVTZ basis set in
    Fig.~\ref{fig:h2o_feast} (Right). The 
    core-excitation is identified by its dominant component in the eigen-vector, corresponding to the 
    transition $1a_1\to 4a_1$ between the orbitals. 


\subsection{CIF-based real-time EOM-CCSD (RT-EOM-CCSD)}
Although the effective Hamiltonian $\bar{H}$ is non-Hermitian, its eigen-values
are real~\footnote{In a truncated space, such as the singles and doubles space
used in EOM-CCSD, the diagonalization can sometimes lead to complex
eigen-values.}, since the similarity transformation preserves the real
eigen-values of the original Hamiltonian. Hence applying the time-evolution
operator $\exp(-\imag\bar{H}t)$ onto a state does not alter its norm. Real-time
electron dynamics within the EOM-CCSD framework has been studied by several
groups~\cite{sonk2011,nascimento2016,nascimento2017,park2019}, based on the
time-dependent Schr{\"o}dinger equation.

In this work, I realize the real-time electron dynamics by applying the
following integral form of the time-evolution operator
\begin{equation}
    \hat{U}(t) = \exp({-\imag\bar{H}t}) = \frac{1}{2\pi \imag}\oint_{C}\frac{e^{-z}}{z\hat{I}-\imag\bar{H}t}\mathrm d{z}.
    \label{equ:U}
\end{equation}
When the contour $C$ encloses the whole spectrum of $\imag\bar{H}t$, this
formulation is exact. In practice, as in many scenarios, one is only interested
in a certain part of the whole spectra. Therefore, one can choose the contour $C$
to define the energy window of interest \footnote{Notice that the eigen-values
of $\imag\bar{H}t$ lie on the imaginary axis. Hence, the contour should be
centered accordingly.}. Using a similar argument as in FEAST-EOM-CCSD, one can
show that the above operator filters the eigen-components whose energies are outside the contour
$C$ and evolves the initial state with the eigen-components whose energies are within the contour $C$ only:
\begin{equation}
    \begin{aligned}
        \ket{\Phi(t)} &= \tilde{{U}}(t)\ket{\Phi(0)} \\
            &= \sum_{p\in \mathcal{C}}e^{-iE_pt}\ket{\Psi^R_p}\braket{\Psi^L_p|\Phi(0)},
    \end{aligned}
     \label{equ:phi_t}
\end{equation}
where I use $\tilde{U}$ to indicate that the time-evolution operator now only
evolves a part of all the eigen-components. Following the steps outlined in
FEAST-EOM-CCSD, to apply $\frac{e^{-z}}{z\hat{I}-\imag \hat{H}\Delta t}$ on the initial
state $\ket{\Phi(t=0)}$ and to carry out the contour integral in an efficient
and accurate way, one can again recast the integral into a weighted summation along a set
of $K$ Gauss-Legendre quadrature nodes on the circle $C$
{
\begin{equation}
    \ket{\Phi(\Delta t)} \approx \frac{1}{2}\sum_{e=1}^{K}\omega_e E_r\Delta t \mathrm e^{\mathrm i \theta_e}\ket{Q_e},
\end{equation}
with the $z_e$ being dependent on  $\Delta t$ to account for the time evolution and 
$\ket{Q_e}$ being the solution of the following linear system problem
\begin{equation}
    (z_e\hat{I}-\imag \hat{H}\Delta t)\ket{Q_e} = e^{-z_e}\ket{\Phi(0)}.
\end{equation}
}
As long as the linear systems are solved accurately and the quadrature errors
are small, in theory, the time evolved state by $\tilde{U}$ suffers from little
to no time step errors. The Gauss-Legendre quadrature nodes ensure that with $K$
nodes within a finite length, a polynomial of degree $2K-1$ can be integrated
exactly along this line. When one chooses the contour $C$ to be away from the
poles of the resolvent, the integrand $e^{-z}/(z\hat{I}-\imag\hat{H}\Delta t)$ is a
smooth function on the contour, and the quadrature error is well controlled by
choosing an appropriate number of $K$ nodes. See Ref.~\cite{takahira2020} for
detailed analysis on the quadrature error. This enables very large
time steps to be used for evolving the system, up to the timescale determined by the fastest
dynamics of interest. 

Currently, I adopt the scheme to update the state repeatedly by the
time-evolution operator with a fixed time step $\Delta t$. In principle,
however, one can also apply the time-evolution operator at different time values
to the initial state to obtain the evolved state at any time. One potential
drawback of this scheme is that the contour will be very large at long times,
and the linear systems at nodes with large real parts will be very difficult to
solve accurately. I leave the investigation along this line to future work.

\subsection{Computational details}
\label{comp}
In the FEAST-EOM-CCSD algorithm, I use $K=10$ Gauss-Legendre quadrature nodes. 
In practice, exploiting the conjugation symmetry of nodes across 
the real axis reduces the number of linear systems from 10 to 5.
However, in the CIF-based
RT-EOM-CCSD algorithm, in order to reduce the quadrature error, $K=16$ Gauss-Legendre 
nodes are used. In the meantime, the poles of $\frac{e^{-z}}{z-\imag \bar{H}\Delta t}$ are on the
imaginary axis, no such conjugation relation can be used. Therefore, one needs to
solve $K$ linear systems for each time step. 
All geometries are optimized at the
CCSD level of theory using the TZVP basis set and can be accessed
at~\cite{cccbdb2024}. 
I use the generalized conjugate residual with inner orthogonalization and outer
truncation (GCROT)~\cite{hicken2010} and a version of 
the generalized minimal residual method (LGMRES)~\cite{baker2005} to solve the linear systems, as
implemented in the SciPy library~\cite{virtanen2020}. 
The latter perfoms better when the linear system is ill-conditioned.

In the FEAST-EOM-CCSD algorithm, the linear systems become ill-conditioned when contour nodes lie close to Hamiltonian eigen-values 
on the real axis. With large radii, only two nodes typically approach eigen-values, 
posing minimal challenges for convergence. However, complications arise with small radii, 
particularly when targeting core-excitation energies surrounded by numerous valence excitations
 in large basis sets 
 and a small fixed number of eigen-states are sought 
 by dynamically reducing the radius of the contour, see Fig~\ref{fig:filter}.
The condition number 
\begin{equation}
\kappa(z_e\hat{I} - \bar{H}) = \frac{\displaystyle\max_{\lambda \in \rm{spec}(\bar{H})} |z_e - \lambda|}{\displaystyle\min_{\lambda \in \rm{spec}(\bar{H})} |z_e - \lambda|} 
\end{equation}
quantifies this difficulty, 
 indicating solution sensitivity to perturbations. 
 When $z_e$ approaches near-degenerate eigen-values, the denominator approaches zero, 
 creating ill-conditioned systems. To address this, I implement a constraint limiting 
 the imaginary component of $z_e$ to a minimum threshold $\delta$ (typically 0.05-0.1 Ha), 
 ensuring a finite denominator in $\kappa$ and well-conditioned linear systems. 
 This technique enables convergence even with near degeneracies present, albeit more slowly
 than in a well-conditioned case, e.g.~using a small basis set when there are not so many valence excitions around 
 a core excitation. 

 Fig.~\ref{fig:filter} illustrates how this constraint affects the energy filtering function, defined as
 a discrete approximation to Eq.~\eqref{equ:cauchy} using 10 Gauss-Legendre quadrature nodes and 
 assuming $f(x)=1$, corresponding to the operator $\bar{P}$ used in FEAST-EOM-CCSD algorithm. 
 It reads
 \begin{equation}
 f(x)\approx \frac{1}{2}\sum_{e=1}^5\mathrm{Re}(\frac{E_r e^{\mathrm{i}\theta_e}\omega_e}{z_e-x}).
 \label{equ:filter}
 \end{equation}
In Fig.~\ref{fig:filter}, the upper panels show where the nodes are located without and with the constraint, respectively;
the lower panels show the energy filtering function Eq.~\eqref{equ:filter} without and with the constraint.
 Without the constraint, the function is sharp, filtering eigen-values outside the energy window defined by $E_c$ and $E_r$.
 With the constraint, the function becomes less sharp but still suppresses eigen-values outside the window,
 allowing convergence to the correct eigen-values. This technique is used to obtain core-excitation 
 energies of molecules in the aug-cc-pVTZ basis set, as shown in Table~\ref{tab:molecules}.

The algorithms are
implemented in PyMES~\cite{liao2021c,liao2023,liaoa}. I use a modified
PySCF~\cite{sun2018a,sun2020} code to perform mean-field HF calculations, to
generate the integrals~\cite{sun2015} and to realize the matrix-vector
multiplication in EOM-CCSD. 
\section{Results and Discussion}
\label{results}
\subsection{Core excitation energies}
\begin{table}[ht!]
    \centering
    \caption{Core excitation ($1a_1\to 4a_1$) energies (eV) of the H$_2$O molecule in various basis sets}
    \begin{tabular}{|c|c|c|c|c|}
        \hline
         Basis & FEAST & ES~$^a$ & EOM-CCSD & Exp.$^b$ \\
        \hline
        6-311G** & 535.74 & 535.72 & 535.76$^c$ & \\
        cc-pVTZ & 535.42 & 535.41 & 535.34$^{d}$ & 534.00 \\
        aug-cc-pVTZ & 535.31 & 535.30 & 535.32$^e$ & \\
        \hline
    \end{tabular}
    \label{tab:h2o}
  
        \footnotesize
        $^a$ Data from Ref.~\cite{peng2015}. $^b$ Data from
        Ref.~\cite{schirmer1993}. $^c$ Data from Ref.~\cite{brabec2012}. $^d$
        Data from Ref.~\cite{sen2013}. $^e$ Data from Ref.~\cite{dutta2014}.
 \end{table}

 \begin{table*}[ht!]
    \centering
        \caption{Core excitation energies (eV) of the N$_2$, CO, C$_2$H$_2$.}
    \begin{tabular}{|c|c|c|c|c|c|c|}
        \hline
         \multirow{2}{*}{Molecule} & \multicolumn{2}{c|}{FEAST} & \multirow{2}{*}{ES~$^a$} & \multirow{2}{*}{CVS~$^b$} & \multirow{2}{*}{EOM~$^c$} &\multirow{2}{*}{Exp.}\\
         \cline{2-3}
         & 6-311G** & aug-cc-pVTZ & & & \\
        \hline
        N$_2$ ($1\sigma_u\to1\pi_g$)  &{402.17} & 401.67 & 401.93 & { 402.04}  & 402.04 &400.96$^d$ \\
        CO ($2\sigma\to2\pi$) & 288.34 & 287.61 & 287.99 & {288.18} & 288.21 & 287.40$^e$ \\
        C$_2$H$_2$ ($1\sigma_u\to1\pi(2p)$) & 286.96 & 286.47 & 286.70 & - & - & 285.81$^f$ \\
        \hline
    \end{tabular}
    \label{tab:molecules}
  
        \footnotesize
        $^a$ Data from Ref.~\cite{peng2015}, with modified Sadlej basis set. $^b$ Data from
        Ref.~\cite{coriani2015}, with aug-cc-pCVTZ+Rydberg basis set. 
        $^c$ Data from Ref.~\cite{coriani2015,coriani2012a} using aug-cc-pCVTZ+Rydberg for N$_2$ and 
        aug-cc-pCVTZ for CO.
        $^d$ Data from Ref.~\cite{myhre2018a}. $^e$ Data
        from Ref.~\cite{domke1990}.  $^f$ Data from Ref.~\cite{ma1991}.
 \end{table*}

\begin{figure*}[ht!]
    \begin{subfigure}{\textwidth}
        \centering
        \includegraphics[width=0.9\textwidth]{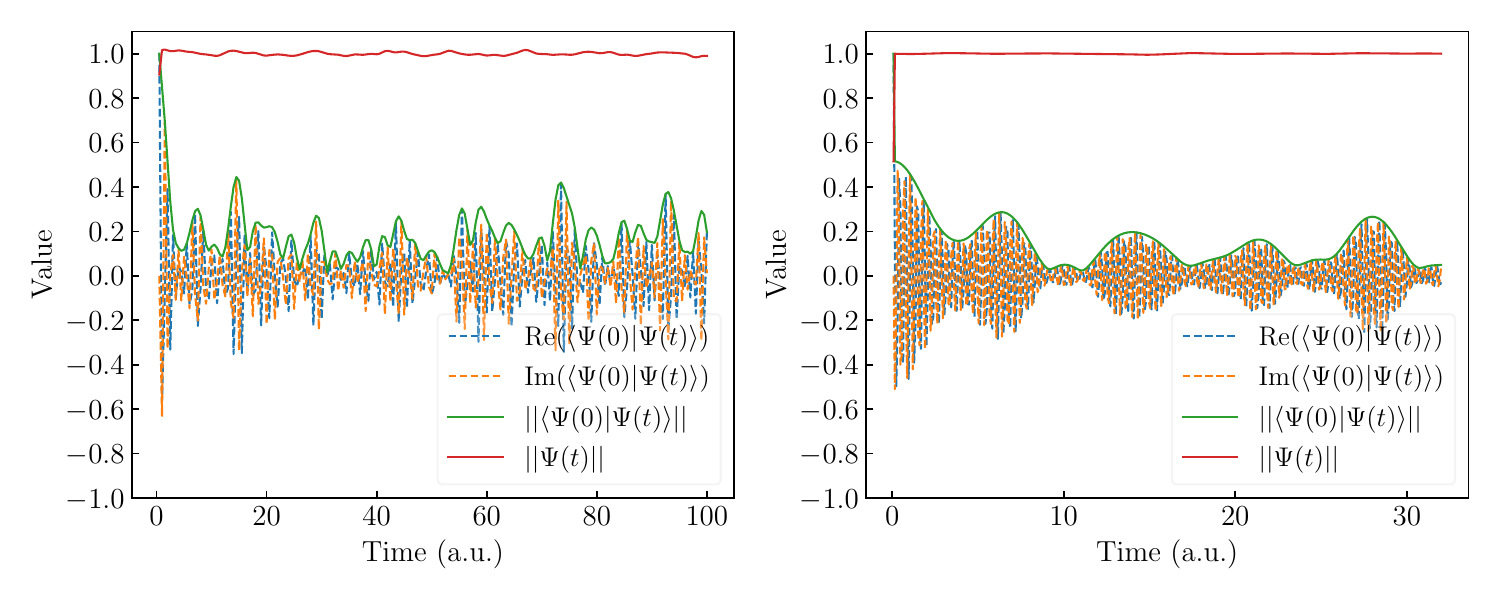}
    \end{subfigure}
    \\
    \begin{subfigure}{\textwidth}
        \centering
        \includegraphics[width=0.9\textwidth]{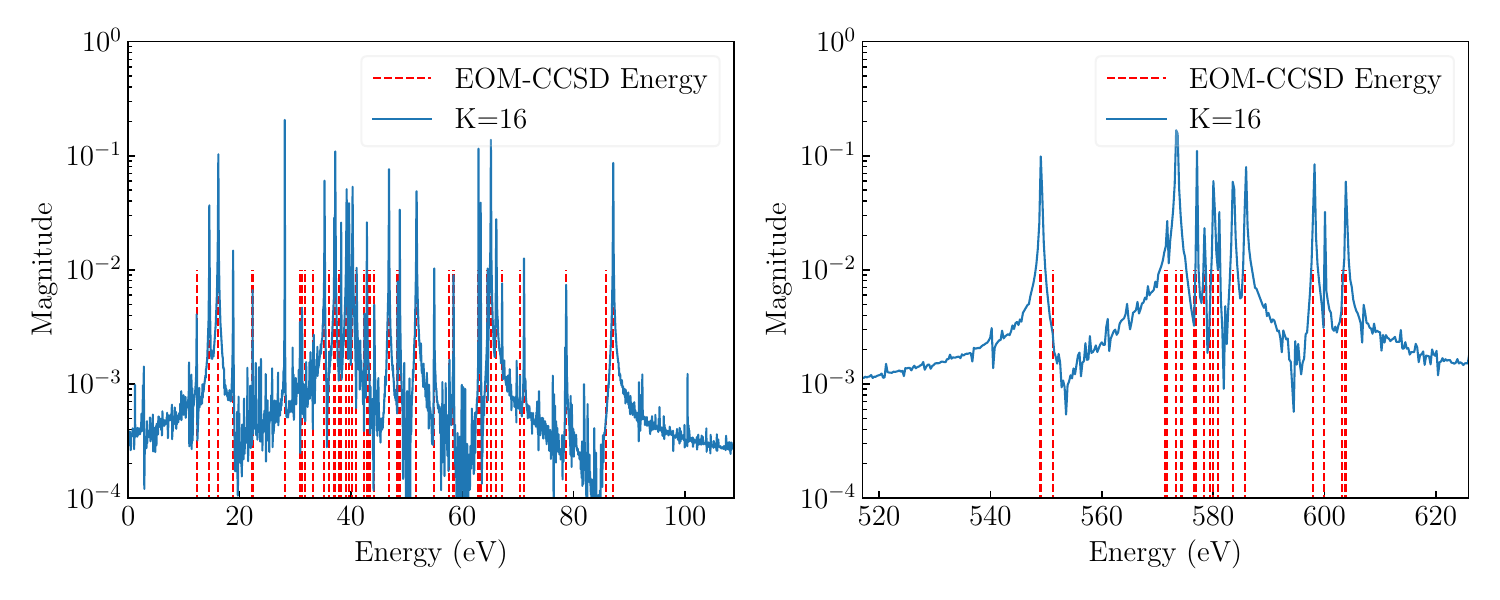}
    \end{subfigure}
    \caption{Upper: The electron dynamics of the H$_2$O molecule in the STO-6G
    basis set with random initial states for valence (left) and core (right) energy regions.
    The time steps used are 0.5~a.u. and
    0.08~a.u. for the valence and core excited states, respectively. In total
    8000 steps are used for both cases (not plotted in full). After each step,
    the state is normalized to 1 to avoid the norm drift. Lower: The Fourier
    transform of the real-time electron dynamics for 
    valence excited state spectra (left) and the core excited
    state spectra (right). The energy windows used for time-evolution are
    -27.211~eV--136.057~eV (-1~Ha--5~Ha) and 272.114~eV--816.342~eV
    (10~Ha--30~Ha) for the valence and core spectra, respectively.  $K=16$ Gauss-Legendre quadrature nodes are used.}
    \label{fig:h2o_td}
\end{figure*}

In Table \ref{tab:h2o}, the oxygen K-edge absorption energies of the water
molecule in various basis sets are calculated using the FEAST-EOM-CCSD
algorithm, compared against those by the energy-specific EOM-CCSD
(ES-EOM-CCSD)~\cite{peng2015}, the canonical
EOM-CCSD~\cite{brabec2012,sen2013,dutta2014} and the experimental
value~\cite{schirmer1993}. In theory, at convergence, the FEAST-EOM-CCSD and the
ES-EOM-CCSD algorithm should agree with each other exactly, because they do not
make additional approximations to the Hamiltonian. The small differences in the
energies between the two algorithms can be attributed to the slightly different
geometries used. 

{ After benchmarking the new algorithm against existing 
methods in the same basis sets, I further apply the FEAST-EOM-CCSD algorithm to
calculate the core excitation energies of the N$_2$, CO, and C$_2$H$_2$ molecules
in the 6-311G** and aug-cc-pVTZ basis sets. The results are shown in
Table~\ref{tab:molecules}. Using the diffused basis set aug-cc-pVTZ, the calculated 
core excitation energies are getting closer to agreement with the experimental values.
Choosing an appropriate basis set that accurately describes orbital relaxation effects is
crucial for predicting core excitation energies. 
Accurately describing orbital relaxation effects through appropriate basis set 
selection is crucial but non-trivial~\cite{carbone2019}.
CVS-EOM-CCSD~\cite{cederbaum1980,coriani2015,vidal2019} is a well-established
method for calculating core excitations efficiently and accurately by 
decoupling the core excitations from the valence excitations. 
Since the CVS-EOM-CCSD and the full EOM-CCSD methods show very good agreement in these cases,
I atrribute the small differences in the energies between FEAST-EOM-CCSD, ES-EOM-CCSD and 
CVS-EOM-CCSD to the different basis sets and geometries used. 
Due to the high density of states around the core excitations 
when a large basis set is used, the convergence of the FEAST-EOM-CCSD algorithm
can be slow. Without the stablization technique described in Computational Details, 
the algorithm may fail to converge. 
This underscores the importance of removing the valence excitation 
space when calculating core excitation energies,
as CVS-EOM-CCSD does. 
Similar to FEAST-EOM-CCSD, ES-EOM-CCSD can also target any specific energy regions 
by selecting trial vectors based on their energies. To converge the calculated energies,
ES-EOM-CCSD requires an increasingly large number of trial vectors to be used, which can be
memory and computationally expensive.
In the future, 
more effcient and accurate algorithms can potentially 
be developed by combining the strengths from
CVS-, ES- and FEAST-EOM-CCSD.}

\subsection{CIF-based real-time electron dynamics}

Due to the many time steps needed for resolving accurately all the energy peaks,
I present only preliminary results of the CIF-based real-time electron dynamics
on the H$_2$O molecule in a minimal basis set. I Fourier transform the overlap
between the initial and evolved states to obtain the spectra and compare the
peaks with the excitation energies computed by the EOM-CCSD. On the upper panel of
Fig.~\ref{fig:h2o_td}, I show the dynamics of two random initial states evolved
for 8000 steps with a time step size of 0.5~a.u and 0.08~a.u. for the valence
(left) and core (right) excited states, respectively. { To put these numbers in 
context, typically a time step size of 0.01~a.u. is 
used for electron dynamics involving core excitations~\cite{nascimento2016,park2019}.}  
I normalize the state
after each time step to avoid the norm drift. The small fluctuations in the norm
of the state testify the quality of the time-evolution operator in the integral
form. Comparatively speaking, the fluctuations in the norm is slightly larger in
the valence excited state case than in the core excited state case. Considering
the extremely large time step size used, 0.5~a.u., this is expected. The energy
windows used for the time-evolutions are -27.211~eV--136.057~eV (-1~Ha--5~Ha)
and 272.114~eV--816.342~eV (10~Ha--30~Ha) for the valence and core spectra,
respectively. The Fourier transform of the overlap between the initial and
evolved state $\braket{\Phi(0)|\Phi(t)}$~\cite{Guther2018} are shown on the
lower panel of Fig.~\ref{fig:h2o_td} for the valence (left) and core (right)
excited states. Most of the energy peaks within the energy windows are resolved
accurately, with a few small valence excitation peaks being obscured by the
noise. This could be due to the fact that these peaks correspond to
eigen-components with very small weights in the initial random state. In
practice, one can always choose an initial state with dominant components in the
energy window of interest to avoid this issue.

\section{Conclusion}
In this work, relying on the operator version of the CIF as the
theoretical foundation for representing holomorphic functions of diagonalizable
operators, I linked the original FEAST algorithm with a much broader class of
algorithms which involve applying functions of diagonalizable operators to a
state. Based on this theoretical insight, two novel algorithms were
proposed--FEAST-EOM-CCSD and CIF-based RT-EOM-CCSD. I benchmarked the
FEAST-EOM-CCSD against EOM-CCSD in calculating low-lying eigen-states in a small
system, and further demonstrated its capability in finding the core excited
states in several molecules, which are relevant to the X-ray spectroscopy. The
results agree with other methods, e.g. ES-EOM-CCSD, CVS-EOM-CCSD and EOM-MRCCSD,
as well as with experiments. { FEAST-EOM-CCSD is highly parallelizable 
at different levels.} For example,
states around different energy centers can be calculated in parallel, 
the Riesz projector can be applied to each trial vector
independently while the underlying linear systems can also be solved
independently. Immediate improvement on the efficiency can be achieved by using
low-scaling approximations~\cite{peng2015} to find better initial trial vectors
and to boost the efficiency of the linear system solver. I believe that the
FEAST-EOM-CCSD algorithm will inspire further developments in the study of
excited states in quantum chemistry and beyond. 

In addition, I reported the first realization of the CIF-based real-time
evolution algorithm  and applied it to the water molecule to target two
energy-separated spectral groups independently with extremely large time step
sizes. The large time step sizes are possible due to the effective
Gauss-Legendre quadrature nodes used for integration. I showed that densely
packed spectra in the valence and core excited regions can be accurately
resolved once long time data of the overlap between the initial and time-evolved
states are gathered. I emphasize that the CIF-based real-time electron dynamics
algorithm is a new approach to realize real-time electron dynamics, and it could
be extended to tensor network theories such as DMRG. Its comparative advantages
and disadvantages in this context over other methods, such as the time-dependent
variational principle~\cite{haegeman2011,haegeman2016,xu2022,yang2020} and
Runge-Kutta~\cite{ronca2017}, are yet to be explored. I envision that the
CIF-based real-time electron dynamics algorithm could also provide insights into
curbing the entanglement growth in tensor network algorithms for simulating the
long time dynamics of quantum many-body systems~\cite{grundner2024}. Its
relationship to other theories that use the complex time for evolution remains
to be understood~\cite{Guther2018,grundner2024}. Finally, other theories and
algorithms that can be cast into the CIF form will also be explored.

\emph{Acknowledgement.} K.L. would like to express gratitude towards Christian
Schilling for his support on this work. Comments from Haibo Ma, Daniel Kats,
Felix Hummel and Ali Alavi are also appreciated. 
Financial support from the
German Research Foundation (Grant SCHI 1476/1-1) is gratefully acknowledged. 

\bibliography{feast}

\providecommand{\latin}[1]{#1}
\makeatletter
\providecommand{\doi}
  {\begingroup\let\do\@makeother\dospecials
  \catcode`\{=1 \catcode`\}=2 \doi@aux}
\providecommand{\doi@aux}[1]{\endgroup\texttt{#1}}
\makeatother
\providecommand*\mcitethebibliography{\thebibliography}
\csname @ifundefined\endcsname{endmcitethebibliography}
  {\let\endmcitethebibliography\endthebibliography}{}
\begin{mcitethebibliography}{85}
\providecommand*\natexlab[1]{#1}
\providecommand*\mciteSetBstSublistMode[1]{}
\providecommand*\mciteSetBstMaxWidthForm[2]{}
\providecommand*\mciteBstWouldAddEndPuncttrue
  {\def\EndOfBibitem{\unskip.}}
\providecommand*\mciteBstWouldAddEndPunctfalse
  {\let\EndOfBibitem\relax}
\providecommand*\mciteSetBstMidEndSepPunct[3]{}
\providecommand*\mciteSetBstSublistLabelBeginEnd[3]{}
\providecommand*\EndOfBibitem{}
\mciteSetBstSublistMode{f}
\mciteSetBstMaxWidthForm{subitem}{(\alph{mcitesubitemcount})}
\mciteSetBstSublistLabelBeginEnd
  {\mcitemaxwidthsubitemform\space}
  {\relax}
  {\relax}

\bibitem[McLACHLAN and BALL(1964)McLACHLAN, and BALL]{mclachlan1964}
McLACHLAN,~A.~D.; BALL,~M.~A. Time-{Dependent} {Hartree}---{Fock} {Theory} for
  {Molecules}. \emph{Reviews of Modern Physics} \textbf{1964}, \emph{36},
  844--855, Publisher: American Physical Society\relax
\mciteBstWouldAddEndPuncttrue
\mciteSetBstMidEndSepPunct{\mcitedefaultmidpunct}
{\mcitedefaultendpunct}{\mcitedefaultseppunct}\relax
\EndOfBibitem
\bibitem[Schönhammer and Gunnarsson(1978)Schönhammer, and
  Gunnarsson]{schonhammer1978}
Schönhammer,~K.; Gunnarsson,~O. Time-dependent approach to the calculation of
  spectral functions. \emph{Physical Review B} \textbf{1978}, \emph{18},
  6606--6614\relax
\mciteBstWouldAddEndPuncttrue
\mciteSetBstMidEndSepPunct{\mcitedefaultmidpunct}
{\mcitedefaultendpunct}{\mcitedefaultseppunct}\relax
\EndOfBibitem
\bibitem[Runge and Gross(1984)Runge, and Gross]{runge1984}
Runge,~E.; Gross,~E. K.~U. Density-{Functional} {Theory} for {Time}-{Dependent}
  {Systems}. \emph{Physical Review Letters} \textbf{1984}, \emph{52},
  997--1000, Publisher: American Physical Society\relax
\mciteBstWouldAddEndPuncttrue
\mciteSetBstMidEndSepPunct{\mcitedefaultmidpunct}
{\mcitedefaultendpunct}{\mcitedefaultseppunct}\relax
\EndOfBibitem
\bibitem[Akama and Nakai(2010)Akama, and Nakai]{akama2010}
Akama,~T.; Nakai,~H. Short-time {Fourier} transform analysis of real-time
  time-dependent {Hartree}–{Fock} and time-dependent density functional
  theory calculations with {Gaussian} basis functions. \emph{The Journal of
  Chemical Physics} \textbf{2010}, \emph{132}, 054104\relax
\mciteBstWouldAddEndPuncttrue
\mciteSetBstMidEndSepPunct{\mcitedefaultmidpunct}
{\mcitedefaultendpunct}{\mcitedefaultseppunct}\relax
\EndOfBibitem
\bibitem[Monkhorst(1977)]{monkhorst1977}
Monkhorst,~H.~J. Calculation of properties with the coupled-cluster method.
  \emph{International Journal of Quantum Chemistry} \textbf{1977}, \emph{12},
  421--432, Publisher: John Wiley \& Sons, Ltd\relax
\mciteBstWouldAddEndPuncttrue
\mciteSetBstMidEndSepPunct{\mcitedefaultmidpunct}
{\mcitedefaultendpunct}{\mcitedefaultseppunct}\relax
\EndOfBibitem
\bibitem[Dalgaard and Monkhorst(1983)Dalgaard, and Monkhorst]{dalgaard1983}
Dalgaard,~E.; Monkhorst,~H.~J. Some aspects of the time-dependent
  coupled-cluster approach to dynamic response functions. \emph{Physical Review
  A} \textbf{1983}, \emph{28}, 1217--1222, Publisher: American Physical
  Society\relax
\mciteBstWouldAddEndPuncttrue
\mciteSetBstMidEndSepPunct{\mcitedefaultmidpunct}
{\mcitedefaultendpunct}{\mcitedefaultseppunct}\relax
\EndOfBibitem
\bibitem[Koch and Jo/rgensen(1990)Koch, and Jo/rgensen]{koch1990}
Koch,~H.; Jo/rgensen,~P. Coupled cluster response functions. \emph{The Journal
  of Chemical Physics} \textbf{1990}, \emph{93}, 3333--3344\relax
\mciteBstWouldAddEndPuncttrue
\mciteSetBstMidEndSepPunct{\mcitedefaultmidpunct}
{\mcitedefaultendpunct}{\mcitedefaultseppunct}\relax
\EndOfBibitem
\bibitem[Stanton and Bartlett(1993)Stanton, and Bartlett]{stanton1993}
Stanton,~J.~F.; Bartlett,~R.~J. The equation of motion coupled‐cluster
  method. {A} systematic biorthogonal approach to molecular excitation
  energies, transition probabilities, and excited state properties. \emph{The
  Journal of Chemical Physics} \textbf{1993}, \emph{98}, 7029--7039\relax
\mciteBstWouldAddEndPuncttrue
\mciteSetBstMidEndSepPunct{\mcitedefaultmidpunct}
{\mcitedefaultendpunct}{\mcitedefaultseppunct}\relax
\EndOfBibitem
\bibitem[Bartlett and Musiał(2007)Bartlett, and Musiał]{bartlett2007b}
Bartlett,~R.~J.; Musiał,~M. Coupled-cluster theory in quantum chemistry.
  \emph{Reviews of Modern Physics} \textbf{2007}, \emph{79}, 291--352,
  Publisher: American Physical Society\relax
\mciteBstWouldAddEndPuncttrue
\mciteSetBstMidEndSepPunct{\mcitedefaultmidpunct}
{\mcitedefaultendpunct}{\mcitedefaultseppunct}\relax
\EndOfBibitem
\bibitem[Krylov(2008)]{krylov2008}
Krylov,~A.~I. Equation-of-{Motion} {Coupled}-{Cluster} {Methods} for
  {Open}-{Shell} and {Electronically} {Excited} {Species}: {The} {Hitchhiker}'s
  {Guide} to {Fock} {Space}. \emph{Annual Review of Physical Chemistry}
  \textbf{2008}, \emph{59}, 433--462, Publisher: Annual Reviews\relax
\mciteBstWouldAddEndPuncttrue
\mciteSetBstMidEndSepPunct{\mcitedefaultmidpunct}
{\mcitedefaultendpunct}{\mcitedefaultseppunct}\relax
\EndOfBibitem
\bibitem[Bartlett(2012)]{bartlett2012}
Bartlett,~R.~J. Coupled-cluster theory and its equation-of-motion extensions.
  \emph{WIREs Computational Molecular Science} \textbf{2012}, \emph{2},
  126--138, \_eprint:
  https://onlinelibrary.wiley.com/doi/pdf/10.1002/wcms.76\relax
\mciteBstWouldAddEndPuncttrue
\mciteSetBstMidEndSepPunct{\mcitedefaultmidpunct}
{\mcitedefaultendpunct}{\mcitedefaultseppunct}\relax
\EndOfBibitem
\bibitem[Sonk \latin{et~al.}(2011)Sonk, Caricato, and Schlegel]{sonk2011}
Sonk,~J.~A.; Caricato,~M.; Schlegel,~H.~B. {TD}-{CI} {Simulation} of the
  {Electronic} {Optical} {Response} of {Molecules} in {Intense} {Fields}:
  {Comparison} of {RPA}, {CIS}, {CIS}({D}), and {EOM}-{CCSD}. \emph{The Journal
  of Physical Chemistry A} \textbf{2011}, \emph{115}, 4678--4690, Publisher:
  American Chemical Society\relax
\mciteBstWouldAddEndPuncttrue
\mciteSetBstMidEndSepPunct{\mcitedefaultmidpunct}
{\mcitedefaultendpunct}{\mcitedefaultseppunct}\relax
\EndOfBibitem
\bibitem[Kvaal(2012)]{kvaal2012}
Kvaal,~S. Ab initio quantum dynamics using coupled-cluster. \emph{The Journal
  of Chemical Physics} \textbf{2012}, \emph{136}, 194109\relax
\mciteBstWouldAddEndPuncttrue
\mciteSetBstMidEndSepPunct{\mcitedefaultmidpunct}
{\mcitedefaultendpunct}{\mcitedefaultseppunct}\relax
\EndOfBibitem
\bibitem[Lopata \latin{et~al.}(2012)Lopata, Van~Kuiken, Khalil, and
  Govind]{lopata2012}
Lopata,~K.; Van~Kuiken,~B.~E.; Khalil,~M.; Govind,~N. Linear-{Response} and
  {Real}-{Time} {Time}-{Dependent} {Density} {Functional} {Theory} {Studies} of
  {Core}-{Level} {Near}-{Edge} {X}-{Ray} {Absorption}. \emph{Journal of
  Chemical Theory and Computation} \textbf{2012}, \emph{8}, 3284--3292,
  Publisher: American Chemical Society\relax
\mciteBstWouldAddEndPuncttrue
\mciteSetBstMidEndSepPunct{\mcitedefaultmidpunct}
{\mcitedefaultendpunct}{\mcitedefaultseppunct}\relax
\EndOfBibitem
\bibitem[Pigg \latin{et~al.}(2012)Pigg, Hagen, Nam, and Papenbrock]{pigg2012a}
Pigg,~D.~A.; Hagen,~G.; Nam,~H.; Papenbrock,~T. Time-dependent coupled-cluster
  method for atomic nuclei. \emph{Physical Review C} \textbf{2012}, \emph{86},
  014308, Publisher: American Physical Society\relax
\mciteBstWouldAddEndPuncttrue
\mciteSetBstMidEndSepPunct{\mcitedefaultmidpunct}
{\mcitedefaultendpunct}{\mcitedefaultseppunct}\relax
\EndOfBibitem
\bibitem[Helgaker \latin{et~al.}(2014)Helgaker, Jørgensen, and
  Olsen]{helgaker2014}
Helgaker,~T.; Jørgensen,~P.; Olsen,~J. \emph{Molecular
  {Electronic}-{Structure} {Theory}}; John Wiley \& Sons, Ltd, 2014; pp
  648--723\relax
\mciteBstWouldAddEndPuncttrue
\mciteSetBstMidEndSepPunct{\mcitedefaultmidpunct}
{\mcitedefaultendpunct}{\mcitedefaultseppunct}\relax
\EndOfBibitem
\bibitem[Nascimento and DePrince(2016)Nascimento, and DePrince]{nascimento2016}
Nascimento,~D.~R.; DePrince,~A.~E. Linear {Absorption} {Spectra} from
  {Explicitly} {Time}-{Dependent} {Equation}-of-{Motion} {Coupled}-{Cluster}
  {Theory}. \emph{Journal of Chemical Theory and Computation} \textbf{2016},
  \emph{12}, 5834--5840\relax
\mciteBstWouldAddEndPuncttrue
\mciteSetBstMidEndSepPunct{\mcitedefaultmidpunct}
{\mcitedefaultendpunct}{\mcitedefaultseppunct}\relax
\EndOfBibitem
\bibitem[Nascimento and DePrince(2017)Nascimento, and DePrince]{nascimento2017}
Nascimento,~D.~R.; DePrince,~A. E.~I. Simulation of {Near}-{Edge} {X}-ray
  {Absorption} {Fine} {Structure} with {Time}-{Dependent}
  {Equation}-of-{Motion} {Coupled}-{Cluster} {Theory}. \emph{The Journal of
  Physical Chemistry Letters} \textbf{2017}, \emph{8}, 2951--2957, Publisher:
  American Chemical Society\relax
\mciteBstWouldAddEndPuncttrue
\mciteSetBstMidEndSepPunct{\mcitedefaultmidpunct}
{\mcitedefaultendpunct}{\mcitedefaultseppunct}\relax
\EndOfBibitem
\bibitem[Pedersen and Kvaal(2019)Pedersen, and Kvaal]{pedersen2019}
Pedersen,~T.~B.; Kvaal,~S. Symplectic integration and physical interpretation
  of time-dependent coupled-cluster theory. \emph{The Journal of Chemical
  Physics} \textbf{2019}, \emph{150}, 144106\relax
\mciteBstWouldAddEndPuncttrue
\mciteSetBstMidEndSepPunct{\mcitedefaultmidpunct}
{\mcitedefaultendpunct}{\mcitedefaultseppunct}\relax
\EndOfBibitem
\bibitem[Kristiansen \latin{et~al.}(2020)Kristiansen,
  Sch\{{\textbackslash}o\}yen, Kvaal, and Pedersen]{kristiansen2020}
Kristiansen,~H.~E.; Sch\{{\textbackslash}o\}yen,~{\textbackslash}.~S.;
  Kvaal,~S.; Pedersen,~T.~B. Numerical stability of time-dependent
  coupled-cluster methods for many-electron dynamics in intense laser pulses.
  \emph{The Journal of Chemical Physics} \textbf{2020}, \emph{152},
  071102\relax
\mciteBstWouldAddEndPuncttrue
\mciteSetBstMidEndSepPunct{\mcitedefaultmidpunct}
{\mcitedefaultendpunct}{\mcitedefaultseppunct}\relax
\EndOfBibitem
\bibitem[Skeidsvoll \latin{et~al.}(2020)Skeidsvoll, Balbi, and
  Koch]{skeidsvoll2020}
Skeidsvoll,~A.~S.; Balbi,~A.; Koch,~H. Time-dependent coupled-cluster theory
  for ultrafast transient-absorption spectroscopy. \emph{Physical Review A}
  \textbf{2020}, \emph{102}, 023115, Publisher: American Physical Society\relax
\mciteBstWouldAddEndPuncttrue
\mciteSetBstMidEndSepPunct{\mcitedefaultmidpunct}
{\mcitedefaultendpunct}{\mcitedefaultseppunct}\relax
\EndOfBibitem
\bibitem[Sverdrup~Ofstad \latin{et~al.}(2023)Sverdrup~Ofstad, Aurbakken,
  Sigmundson~Sch\{{\textbackslash}o\}yen, Kristiansen, Kvaal, and
  Pedersen]{sverdrupofstad2023}
Sverdrup~Ofstad,~B.; Aurbakken,~E.;
  Sigmundson~Sch\{{\textbackslash}o\}yen,~{\textbackslash}.;
  Kristiansen,~H.~E.; Kvaal,~S.; Pedersen,~T.~B. Time-dependent coupled-cluster
  theory. \emph{WIREs Computational Molecular Science} \textbf{2023},
  \emph{13}, e1666, \_eprint:
  https://onlinelibrary.wiley.com/doi/pdf/10.1002/wcms.1666\relax
\mciteBstWouldAddEndPuncttrue
\mciteSetBstMidEndSepPunct{\mcitedefaultmidpunct}
{\mcitedefaultendpunct}{\mcitedefaultseppunct}\relax
\EndOfBibitem
\bibitem[Cederbaum \latin{et~al.}(1980)Cederbaum, Domcke, and
  Schirmer]{cederbaum1980}
Cederbaum,~L.~S.; Domcke,~W.; Schirmer,~J. Many-body theory of core holes.
  \emph{Physical Review A} \textbf{1980}, \emph{22}, 206--222, Publisher:
  American Physical Society\relax
\mciteBstWouldAddEndPuncttrue
\mciteSetBstMidEndSepPunct{\mcitedefaultmidpunct}
{\mcitedefaultendpunct}{\mcitedefaultseppunct}\relax
\EndOfBibitem
\bibitem[Coriani and Koch(2015)Coriani, and Koch]{coriani2015}
Coriani,~S.; Koch,~H. Communication: {X}-ray absorption spectra and
  core-ionization potentials within a core-valence separated coupled cluster
  framework. \emph{The Journal of Chemical Physics} \textbf{2015}, \emph{143},
  181103\relax
\mciteBstWouldAddEndPuncttrue
\mciteSetBstMidEndSepPunct{\mcitedefaultmidpunct}
{\mcitedefaultendpunct}{\mcitedefaultseppunct}\relax
\EndOfBibitem
\bibitem[Vidal \latin{et~al.}(2019)Vidal, Feng, Epifanovsky, Krylov, and
  Coriani]{vidal2019}
Vidal,~M.~L.; Feng,~X.; Epifanovsky,~E.; Krylov,~A.~I.; Coriani,~S. New and
  {Efficient} {Equation}-of-{Motion} {Coupled}-{Cluster} {Framework} for
  {Core}-{Excited} and {Core}-{Ionized} {States}. \emph{Journal of Chemical
  Theory and Computation} \textbf{2019}, \emph{15}, 3117--3133, Publisher:
  American Chemical Society\relax
\mciteBstWouldAddEndPuncttrue
\mciteSetBstMidEndSepPunct{\mcitedefaultmidpunct}
{\mcitedefaultendpunct}{\mcitedefaultseppunct}\relax
\EndOfBibitem
\bibitem[Peng \latin{et~al.}(2015)Peng, Lestrange, Goings, Caricato, and
  Li]{peng2015}
Peng,~B.; Lestrange,~P.~J.; Goings,~J.~J.; Caricato,~M.; Li,~X.
  Energy-{Specific} {Equation}-of-{Motion} {Coupled}-{Cluster} {Methods} for
  {High}-{Energy} {Excited} {States}: {Application} to {K}-edge {X}-ray
  {Absorption} {Spectroscopy}. \emph{Journal of Chemical Theory and
  Computation} \textbf{2015}, \emph{11}, 4146--4153, Publisher: American
  Chemical Society\relax
\mciteBstWouldAddEndPuncttrue
\mciteSetBstMidEndSepPunct{\mcitedefaultmidpunct}
{\mcitedefaultendpunct}{\mcitedefaultseppunct}\relax
\EndOfBibitem
\bibitem[Yu \latin{et~al.}(2017)Yu, Pekker, and Clark]{yu2017}
Yu,~X.; Pekker,~D.; Clark,~B.~K. Finding {Matrix} {Product} {State}
  {Representations} of {Highly} {Excited} {Eigenstates} of {Many}-{Body}
  {Localized} {Hamiltonians}. \emph{Physical Review Letters} \textbf{2017},
  \emph{118}, 017201, Publisher: American Physical Society\relax
\mciteBstWouldAddEndPuncttrue
\mciteSetBstMidEndSepPunct{\mcitedefaultmidpunct}
{\mcitedefaultendpunct}{\mcitedefaultseppunct}\relax
\EndOfBibitem
\bibitem[Dorando \latin{et~al.}(2007)Dorando, Hachmann, and Chan]{dorando2007a}
Dorando,~J.~J.; Hachmann,~J.; Chan,~G. K.-L. Targeted excited state algorithms.
  \emph{The Journal of Chemical Physics} \textbf{2007}, \emph{127},
  084109\relax
\mciteBstWouldAddEndPuncttrue
\mciteSetBstMidEndSepPunct{\mcitedefaultmidpunct}
{\mcitedefaultendpunct}{\mcitedefaultseppunct}\relax
\EndOfBibitem
\bibitem[Booth and Chan(2012)Booth, and Chan]{booth2012}
Booth,~G.~H.; Chan,~G. K.-L. Communication: {Excited} states, dynamic
  correlation functions and spectral properties from full configuration
  interaction quantum {Monte} {Carlo}. \emph{The Journal of Chemical Physics}
  \textbf{2012}, \emph{137}, 191102\relax
\mciteBstWouldAddEndPuncttrue
\mciteSetBstMidEndSepPunct{\mcitedefaultmidpunct}
{\mcitedefaultendpunct}{\mcitedefaultseppunct}\relax
\EndOfBibitem
\bibitem[Zhao and Neuscamman(2019)Zhao, and Neuscamman]{zhao2019}
Zhao,~L.; Neuscamman,~E. Variational {Excitations} in {Real} {Solids}:
  {Optical} {Gaps} and {Insights} into {Many}-{Body} {Perturbation} {Theory}.
  \emph{Physical Review Letters} \textbf{2019}, \emph{123}, 036402\relax
\mciteBstWouldAddEndPuncttrue
\mciteSetBstMidEndSepPunct{\mcitedefaultmidpunct}
{\mcitedefaultendpunct}{\mcitedefaultseppunct}\relax
\EndOfBibitem
\bibitem[Kühner and White(1999)Kühner, and White]{kuhner1999}
Kühner,~T.~D.; White,~S.~R. Dynamical correlation functions using the density
  matrix renormalization group. \emph{Physical Review B} \textbf{1999},
  \emph{60}, 335--343, Publisher: American Physical Society\relax
\mciteBstWouldAddEndPuncttrue
\mciteSetBstMidEndSepPunct{\mcitedefaultmidpunct}
{\mcitedefaultendpunct}{\mcitedefaultseppunct}\relax
\EndOfBibitem
\bibitem[McClain \latin{et~al.}(2016)McClain, Lischner, Watson, Matthews,
  Ronca, Louie, Berkelbach, and Chan]{mcclain2016b}
McClain,~J.; Lischner,~J.; Watson,~T.; Matthews,~D.~A.; Ronca,~E.;
  Louie,~S.~G.; Berkelbach,~T.~C.; Chan,~G. K.-L. Spectral functions of the
  uniform electron gas via coupled-cluster theory and comparison to the
  \${GW}\$ and related approximations. \emph{Physical Review B} \textbf{2016},
  \emph{93}, 235139, Publisher: American Physical Society\relax
\mciteBstWouldAddEndPuncttrue
\mciteSetBstMidEndSepPunct{\mcitedefaultmidpunct}
{\mcitedefaultendpunct}{\mcitedefaultseppunct}\relax
\EndOfBibitem
\bibitem[Lewis and Berkelbach(2019)Lewis, and Berkelbach]{lewis2019a}
Lewis,~A.~M.; Berkelbach,~T.~C. Ab {Initio} {Lifetime} and {Concomitant}
  {Double}-{Excitation} {Character} of {Plasmons} at {Metallic} {Densities}.
  \emph{Physical Review Letters} \textbf{2019}, \emph{122}, 226402, Publisher:
  American Physical Society\relax
\mciteBstWouldAddEndPuncttrue
\mciteSetBstMidEndSepPunct{\mcitedefaultmidpunct}
{\mcitedefaultendpunct}{\mcitedefaultseppunct}\relax
\EndOfBibitem
\bibitem[Lee \latin{et~al.}(2023)Lee, Zhai, and Chan]{lee2023b}
Lee,~S.; Zhai,~H.; Chan,~G. K.-L. An {Ab} {Initio} {Correction} {Vector}
  {Restricted} {Active} {Space} {Approach} to the {L}-{Edge} {XAS} and 2p3d
  {RIXS} {Spectra} of {Transition} {Metal} {Complexes}. \emph{Journal of
  Chemical Theory and Computation} \textbf{2023}, \emph{19}, 7753--7763,
  Publisher: American Chemical Society\relax
\mciteBstWouldAddEndPuncttrue
\mciteSetBstMidEndSepPunct{\mcitedefaultmidpunct}
{\mcitedefaultendpunct}{\mcitedefaultseppunct}\relax
\EndOfBibitem
\bibitem[Kristensen \latin{et~al.}(2009)Kristensen, Kauczor, Kjærgaard, and
  Jørgensen]{kristensen2009}
Kristensen,~K.; Kauczor,~J.; Kjærgaard,~T.; Jørgensen,~P. Quasienergy
  formulation of damped response theory. \emph{The Journal of Chemical Physics}
  \textbf{2009}, \emph{131}, 044112\relax
\mciteBstWouldAddEndPuncttrue
\mciteSetBstMidEndSepPunct{\mcitedefaultmidpunct}
{\mcitedefaultendpunct}{\mcitedefaultseppunct}\relax
\EndOfBibitem
\bibitem[Schnack-Petersen \latin{et~al.}(2023)Schnack-Petersen, Moitra,
  Folkestad, and Coriani]{schnack-petersen2023}
Schnack-Petersen,~A.~K.; Moitra,~T.; Folkestad,~S.~D.; Coriani,~S. New
  {Implementation} of an {Equation}-of-{Motion} {Coupled}-{Cluster}
  {Damped}-{Response} {Framework} with {Illustrative} {Applications} to
  {Resonant} {Inelastic} {X}-ray {Scattering}. \emph{The Journal of Physical
  Chemistry A} \textbf{2023}, \emph{127}, 1775--1793, Publisher: American
  Chemical Society\relax
\mciteBstWouldAddEndPuncttrue
\mciteSetBstMidEndSepPunct{\mcitedefaultmidpunct}
{\mcitedefaultendpunct}{\mcitedefaultseppunct}\relax
\EndOfBibitem
\bibitem[Norman \latin{et~al.}(2001)Norman, Bishop, Jo/rgen Aa.~Jensen, and
  Oddershede]{norman2001}
Norman,~P.; Bishop,~D.~M.; Jo/rgen Aa.~Jensen,~H.; Oddershede,~J. Near-resonant
  absorption in the time-dependent self-consistent field and
  multiconfigurational self-consistent field approximations. \emph{The Journal
  of Chemical Physics} \textbf{2001}, \emph{115}, 10323--10334\relax
\mciteBstWouldAddEndPuncttrue
\mciteSetBstMidEndSepPunct{\mcitedefaultmidpunct}
{\mcitedefaultendpunct}{\mcitedefaultseppunct}\relax
\EndOfBibitem
\bibitem[Norman \latin{et~al.}(2005)Norman, Bishop, Jensen, and
  Oddershede]{norman2005}
Norman,~P.; Bishop,~D.~M.; Jensen,~H. J.~A.; Oddershede,~J. Nonlinear response
  theory with relaxation: {The} first-order hyperpolarizability. \emph{The
  Journal of Chemical Physics} \textbf{2005}, \emph{123}, 194103\relax
\mciteBstWouldAddEndPuncttrue
\mciteSetBstMidEndSepPunct{\mcitedefaultmidpunct}
{\mcitedefaultendpunct}{\mcitedefaultseppunct}\relax
\EndOfBibitem
\bibitem[Ekström \latin{et~al.}(2006)Ekström, Norman, Carravetta, and
  Ågren]{ekstrom2006}
Ekström,~U.; Norman,~P.; Carravetta,~V.; Ågren,~H. Polarization {Propagator}
  for {X}-{Ray} {Spectra}. \emph{Physical Review Letters} \textbf{2006},
  \emph{97}, 143001, Publisher: American Physical Society\relax
\mciteBstWouldAddEndPuncttrue
\mciteSetBstMidEndSepPunct{\mcitedefaultmidpunct}
{\mcitedefaultendpunct}{\mcitedefaultseppunct}\relax
\EndOfBibitem
\bibitem[Coriani \latin{et~al.}(2012)Coriani, Fransson, Christiansen, and
  Norman]{coriani2012}
Coriani,~S.; Fransson,~T.; Christiansen,~O.; Norman,~P.
  Asymmetric-{Lanczos}-{Chain}-{Driven} {Implementation} of {Electronic}
  {Resonance} {Convergent} {Coupled}-{Cluster} {Linear} {Response} {Theory}.
  \emph{Journal of Chemical Theory and Computation} \textbf{2012}, \emph{8},
  1616--1628, Publisher: American Chemical Society\relax
\mciteBstWouldAddEndPuncttrue
\mciteSetBstMidEndSepPunct{\mcitedefaultmidpunct}
{\mcitedefaultendpunct}{\mcitedefaultseppunct}\relax
\EndOfBibitem
\bibitem[Coriani \latin{et~al.}(2012)Coriani, Christiansen, Fransson, and
  Norman]{coriani2012a}
Coriani,~S.; Christiansen,~O.; Fransson,~T.; Norman,~P. Coupled-cluster
  response theory for near-edge x-ray-absorption fine structure of atoms and
  molecules. \emph{Physical Review A} \textbf{2012}, \emph{85}, 022507,
  Publisher: American Physical Society\relax
\mciteBstWouldAddEndPuncttrue
\mciteSetBstMidEndSepPunct{\mcitedefaultmidpunct}
{\mcitedefaultendpunct}{\mcitedefaultseppunct}\relax
\EndOfBibitem
\bibitem[Pedersen \latin{et~al.}(2014)Pedersen, Hedegård, Olsen, Kauczor,
  Norman, and Kongsted]{pedersen2014}
Pedersen,~M.~N.; Hedegård,~E.~D.; Olsen,~J. M.~H.; Kauczor,~J.; Norman,~P.;
  Kongsted,~J. Damped {Response} {Theory} in {Combination} with {Polarizable}
  {Environments}: {The} {Polarizable} {Embedding} {Complex} {Polarization}
  {Propagator} {Method}. \emph{Journal of Chemical Theory and Computation}
  \textbf{2014}, \emph{10}, 1164--1171, Publisher: American Chemical
  Society\relax
\mciteBstWouldAddEndPuncttrue
\mciteSetBstMidEndSepPunct{\mcitedefaultmidpunct}
{\mcitedefaultendpunct}{\mcitedefaultseppunct}\relax
\EndOfBibitem
\bibitem[Riesz and Sz.-Nagy(2012)Riesz, and Sz.-Nagy]{riesz2012}
Riesz,~F.; Sz.-Nagy,~B. \emph{Functional {Analysis}}; Courier Corporation,
  2012; Google-Books-ID: JCfEAgAAQBAJ\relax
\mciteBstWouldAddEndPuncttrue
\mciteSetBstMidEndSepPunct{\mcitedefaultmidpunct}
{\mcitedefaultendpunct}{\mcitedefaultseppunct}\relax
\EndOfBibitem
\bibitem[Polizzi(2009)]{polizzi2009}
Polizzi,~E. Density-matrix-based algorithm for solving eigenvalue problems.
  \emph{Physical Review B} \textbf{2009}, \emph{79}, 115112\relax
\mciteBstWouldAddEndPuncttrue
\mciteSetBstMidEndSepPunct{\mcitedefaultmidpunct}
{\mcitedefaultendpunct}{\mcitedefaultseppunct}\relax
\EndOfBibitem
\bibitem[Di~Napoli \latin{et~al.}(2016)Di~Napoli, Polizzi, and
  Saad]{dinapoli2016}
Di~Napoli,~E.; Polizzi,~E.; Saad,~Y. Efficient estimation of eigenvalue counts
  in an interval. \emph{Numerical Linear Algebra with Applications}
  \textbf{2016}, \emph{23}, 674--692, \_eprint:
  https://onlinelibrary.wiley.com/doi/pdf/10.1002/nla.2048\relax
\mciteBstWouldAddEndPuncttrue
\mciteSetBstMidEndSepPunct{\mcitedefaultmidpunct}
{\mcitedefaultendpunct}{\mcitedefaultseppunct}\relax
\EndOfBibitem
\bibitem[Baiardi \latin{et~al.}(2022)Baiardi, Kelemen, and
  Reiher]{baiardi2022b}
Baiardi,~A.; Kelemen,~A.~K.; Reiher,~M. Excited-{State} {DMRG} {Made} {Simple}
  with {FEAST}. \emph{Journal of Chemical Theory and Computation}
  \textbf{2022}, \emph{18}, 415--430, Publisher: American Chemical
  Society\relax
\mciteBstWouldAddEndPuncttrue
\mciteSetBstMidEndSepPunct{\mcitedefaultmidpunct}
{\mcitedefaultendpunct}{\mcitedefaultseppunct}\relax
\EndOfBibitem
\bibitem[Takahira \latin{et~al.}(2020)Takahira, Ohashi, Sogabe, and
  Usuda]{takahira2020}
Takahira,~S.; Ohashi,~A.; Sogabe,~T.; Usuda,~T.~S. Quantum algorithm for matrix
  functions by {Cauchy}'s integral formula. \emph{Quantum Information and
  Computation} \textbf{2020}, \emph{20}, 14--36, arXiv:2106.08075
  [quant-ph]\relax
\mciteBstWouldAddEndPuncttrue
\mciteSetBstMidEndSepPunct{\mcitedefaultmidpunct}
{\mcitedefaultendpunct}{\mcitedefaultseppunct}\relax
\EndOfBibitem
\bibitem[Hummel(2018)]{hummel2018}
Hummel,~F. Finite {Temperature} {Coupled} {Cluster} {Theories} for {Extended}
  {Systems}. \emph{Journal of Chemical Theory and Computation} \textbf{2018},
  \emph{14}, 6505--6514, Publisher: American Chemical Society\relax
\mciteBstWouldAddEndPuncttrue
\mciteSetBstMidEndSepPunct{\mcitedefaultmidpunct}
{\mcitedefaultendpunct}{\mcitedefaultseppunct}\relax
\EndOfBibitem
\bibitem[White and Chan(2018)White, and Chan]{white2018}
White,~A.~F.; Chan,~G. K.-L. A {Time}-{Dependent} {Formulation} of
  {Coupled}-{Cluster} {Theory} for {Many}-{Fermion} {Systems} at {Finite}
  {Temperature}. \emph{Journal of Chemical Theory and Computation}
  \textbf{2018}, \emph{14}, 5690--5700, Publisher: American Chemical
  Society\relax
\mciteBstWouldAddEndPuncttrue
\mciteSetBstMidEndSepPunct{\mcitedefaultmidpunct}
{\mcitedefaultendpunct}{\mcitedefaultseppunct}\relax
\EndOfBibitem
\bibitem[Harsha \latin{et~al.}(2019)Harsha, Henderson, and
  Scuseria]{harsha2019}
Harsha,~G.; Henderson,~T.~M.; Scuseria,~G.~E. Thermofield {Theory} for
  {Finite}-{Temperature} {Coupled} {Cluster}. \emph{Journal of Chemical Theory
  and Computation} \textbf{2019}, \emph{15}, 6127--6136, Publisher: American
  Chemical Society\relax
\mciteBstWouldAddEndPuncttrue
\mciteSetBstMidEndSepPunct{\mcitedefaultmidpunct}
{\mcitedefaultendpunct}{\mcitedefaultseppunct}\relax
\EndOfBibitem
\bibitem[Deutsch(1991)]{deutsch1991}
Deutsch,~J.~M. Quantum statistical mechanics in a closed system. \emph{Physical
  Review A} \textbf{1991}, \emph{43}, 2046--2049, Publisher: American Physical
  Society\relax
\mciteBstWouldAddEndPuncttrue
\mciteSetBstMidEndSepPunct{\mcitedefaultmidpunct}
{\mcitedefaultendpunct}{\mcitedefaultseppunct}\relax
\EndOfBibitem
\bibitem[Srednicki(1994)]{srednicki1994}
Srednicki,~M. Chaos and quantum thermalization. \emph{Physical Review E}
  \textbf{1994}, \emph{50}, 888--901, Publisher: American Physical
  Society\relax
\mciteBstWouldAddEndPuncttrue
\mciteSetBstMidEndSepPunct{\mcitedefaultmidpunct}
{\mcitedefaultendpunct}{\mcitedefaultseppunct}\relax
\EndOfBibitem
\bibitem[Rigol \latin{et~al.}(2008)Rigol, Dunjko, and Olshanii]{rigol2008}
Rigol,~M.; Dunjko,~V.; Olshanii,~M. Thermalization and its mechanism for
  generic isolated quantum systems. \emph{Nature} \textbf{2008}, \emph{452},
  854--858\relax
\mciteBstWouldAddEndPuncttrue
\mciteSetBstMidEndSepPunct{\mcitedefaultmidpunct}
{\mcitedefaultendpunct}{\mcitedefaultseppunct}\relax
\EndOfBibitem
\bibitem[Deutsch(2018)]{deutsch2018}
Deutsch,~J.~M. Eigenstate thermalization hypothesis. \emph{Reports on Progress
  in Physics} \textbf{2018}, \emph{81}, 082001, Publisher: IOP Publishing\relax
\mciteBstWouldAddEndPuncttrue
\mciteSetBstMidEndSepPunct{\mcitedefaultmidpunct}
{\mcitedefaultendpunct}{\mcitedefaultseppunct}\relax
\EndOfBibitem
\bibitem[Irmejs \latin{et~al.}(2024)Irmejs, Bañuls, and Cirac]{irmejs2024}
Irmejs,~R.; Bañuls,~M.~C.; Cirac,~J.~I. Efficient {Quantum} {Algorithm} for
  {Filtering} {Product} {States}. \emph{Quantum} \textbf{2024}, \emph{8}, 1389,
  arXiv:2312.13892 [quant-ph]\relax
\mciteBstWouldAddEndPuncttrue
\mciteSetBstMidEndSepPunct{\mcitedefaultmidpunct}
{\mcitedefaultendpunct}{\mcitedefaultseppunct}\relax
\EndOfBibitem
\bibitem[Emrich(1981)]{emrich1981}
Emrich,~K. An extension of the coupled cluster formalism to excited states
  ({I}). \emph{Nuclear Physics A} \textbf{1981}, \emph{351}, 379--396\relax
\mciteBstWouldAddEndPuncttrue
\mciteSetBstMidEndSepPunct{\mcitedefaultmidpunct}
{\mcitedefaultendpunct}{\mcitedefaultseppunct}\relax
\EndOfBibitem
\bibitem[Emrich(1981)]{emrich1981a}
Emrich,~K. An extension of the coupled cluster formalism to excited states:
  ({II}). {Approximations} and tests. \emph{Nuclear Physics A} \textbf{1981},
  \emph{351}, 397--438\relax
\mciteBstWouldAddEndPuncttrue
\mciteSetBstMidEndSepPunct{\mcitedefaultmidpunct}
{\mcitedefaultendpunct}{\mcitedefaultseppunct}\relax
\EndOfBibitem
\bibitem[Harris \latin{et~al.}(2020)Harris, Millman, van~der Walt, Gommers,
  Virtanen, Cournapeau, Wieser, Taylor, Berg, Smith, Kern, Picus, Hoyer, van
  Kerkwijk, Brett, Haldane, del Río, Wiebe, Peterson, Gérard-Marchant,
  Sheppard, Reddy, Weckesser, Abbasi, Gohlke, and Oliphant]{Harris2020}
Harris,~C.~R. \latin{et~al.}  Array programming with {NumPy}. \emph{Nature}
  \textbf{2020}, \emph{585}, 357--362, Number: 7825 Publisher: Nature
  Publishing Group\relax
\mciteBstWouldAddEndPuncttrue
\mciteSetBstMidEndSepPunct{\mcitedefaultmidpunct}
{\mcitedefaultendpunct}{\mcitedefaultseppunct}\relax
\EndOfBibitem
\bibitem[Park \latin{et~al.}(2019)Park, Perera, and Bartlett]{park2019}
Park,~Y.~C.; Perera,~A.; Bartlett,~R.~J. Equation of motion coupled-cluster for
  core excitation spectra: {Two} complementary approaches. \emph{The Journal of
  Chemical Physics} \textbf{2019}, \emph{151}, 164117\relax
\mciteBstWouldAddEndPuncttrue
\mciteSetBstMidEndSepPunct{\mcitedefaultmidpunct}
{\mcitedefaultendpunct}{\mcitedefaultseppunct}\relax
\EndOfBibitem
\bibitem[ccc()]{cccbdb2024}
{CCCBDB} {List} calculated geometries, accessed on {Aug} 30, 2024.
  \url{https://cccbdb.nist.gov/geom1x.asp}\relax
\mciteBstWouldAddEndPuncttrue
\mciteSetBstMidEndSepPunct{\mcitedefaultmidpunct}
{\mcitedefaultendpunct}{\mcitedefaultseppunct}\relax
\EndOfBibitem
\bibitem[Hicken and Zingg(2010)Hicken, and Zingg]{hicken2010}
Hicken,~J.~E.; Zingg,~D.~W. A {Simplified} and {Flexible} {Variant} of {GCROT}
  for {Solving} {Nonsymmetric} {Linear} {Systems}. \emph{SIAM Journal on
  Scientific Computing} \textbf{2010}, \emph{32}, 1672--1694, Publisher:
  Society for Industrial and Applied Mathematics\relax
\mciteBstWouldAddEndPuncttrue
\mciteSetBstMidEndSepPunct{\mcitedefaultmidpunct}
{\mcitedefaultendpunct}{\mcitedefaultseppunct}\relax
\EndOfBibitem
\bibitem[Baker \latin{et~al.}(2005)Baker, Jessup, and Manteuffel]{baker2005}
Baker,~A.~H.; Jessup,~E.~R.; Manteuffel,~T. A {Technique} for {Accelerating}
  the {Convergence} of {Restarted} {GMRES}. \emph{SIAM Journal on Matrix
  Analysis and Applications} \textbf{2005}, \emph{26}, 962--984, Publisher:
  Society for Industrial and Applied Mathematics\relax
\mciteBstWouldAddEndPuncttrue
\mciteSetBstMidEndSepPunct{\mcitedefaultmidpunct}
{\mcitedefaultendpunct}{\mcitedefaultseppunct}\relax
\EndOfBibitem
\bibitem[Virtanen \latin{et~al.}(2020)Virtanen, Gommers, Oliphant, Haberland,
  Reddy, Cournapeau, Burovski, Peterson, Weckesser, Bright, van~der Walt,
  Brett, Wilson, Millman, Mayorov, Nelson, Jones, Kern, Larson, Carey, Polat,
  Feng, Moore, VanderPlas, Laxalde, Perktold, Cimrman, Henriksen, Quintero,
  Harris, Archibald, Ribeiro, Pedregosa, and van Mulbregt]{virtanen2020}
Virtanen,~P. \latin{et~al.}  {SciPy} 1.0: fundamental algorithms for scientific
  computing in {Python}. \emph{Nature Methods} \textbf{2020}, \emph{17},
  261--272, Publisher: Nature Publishing Group\relax
\mciteBstWouldAddEndPuncttrue
\mciteSetBstMidEndSepPunct{\mcitedefaultmidpunct}
{\mcitedefaultendpunct}{\mcitedefaultseppunct}\relax
\EndOfBibitem
\bibitem[Liao \latin{et~al.}(2021)Liao, Schraivogel, Luo, Kats, and
  Alavi]{liao2021c}
Liao,~K.; Schraivogel,~T.; Luo,~H.; Kats,~D.; Alavi,~A. Towards efficient and
  accurate ab initio solutions to periodic systems via transcorrelation and
  coupled cluster theory. \emph{Physical Review Research} \textbf{2021},
  \emph{3}, 033072, Publisher: American Physical Society\relax
\mciteBstWouldAddEndPuncttrue
\mciteSetBstMidEndSepPunct{\mcitedefaultmidpunct}
{\mcitedefaultendpunct}{\mcitedefaultseppunct}\relax
\EndOfBibitem
\bibitem[Liao \latin{et~al.}(2023)Liao, Zhai, Christlmaier, Schraivogel, Ríos,
  Kats, and Alavi]{liao2023}
Liao,~K.; Zhai,~H.; Christlmaier,~E.~M.; Schraivogel,~T.; Ríos,~P.~L.;
  Kats,~D.; Alavi,~A. Density {Matrix} {Renormalization} {Group} for
  {Transcorrelated} {Hamiltonians}: {Ground} and {Excited} {States} in
  {Molecules}. \emph{Journal of Chemical Theory and Computation} \textbf{2023},
  \emph{19}, 1734--1743, Publisher: American Chemical Society\relax
\mciteBstWouldAddEndPuncttrue
\mciteSetBstMidEndSepPunct{\mcitedefaultmidpunct}
{\mcitedefaultendpunct}{\mcitedefaultseppunct}\relax
\EndOfBibitem
\bibitem[Liao()]{liaoa}
Liao,~K. nickirk/pymes at main.
  \url{https://github.com/nickirk/pymes/tree/main}\relax
\mciteBstWouldAddEndPuncttrue
\mciteSetBstMidEndSepPunct{\mcitedefaultmidpunct}
{\mcitedefaultendpunct}{\mcitedefaultseppunct}\relax
\EndOfBibitem
\bibitem[Sun \latin{et~al.}(2018)Sun, Berkelbach, Blunt, Booth, Guo, Li, Liu,
  McClain, Sayfutyarova, Sharma, Wouters, and Chan]{sun2018a}
Sun,~Q.; Berkelbach,~T.~C.; Blunt,~N.~S.; Booth,~G.~H.; Guo,~S.; Li,~Z.;
  Liu,~J.; McClain,~J.~D.; Sayfutyarova,~E.~R.; Sharma,~S.; Wouters,~S.;
  Chan,~G. K.-L. {PySCF}: the {Python}‐based simulations of chemistry
  framework. \emph{Wiley Interdisciplinary Reviews: Computational Molecular
  Science} \textbf{2018}, \emph{8}, Publisher: John Wiley \& Sons, Ltd\relax
\mciteBstWouldAddEndPuncttrue
\mciteSetBstMidEndSepPunct{\mcitedefaultmidpunct}
{\mcitedefaultendpunct}{\mcitedefaultseppunct}\relax
\EndOfBibitem
\bibitem[Sun \latin{et~al.}(2020)Sun, Zhang, Banerjee, Bao, Barbry, Blunt,
  Bogdanov, Booth, Chen, Cui, Eriksen, Gao, Guo, Hermann, Hermes, Koh, Koval,
  Lehtola, Li, Liu, Mardirossian, McClain, Motta, Mussard, Pham, Pulkin,
  Purwanto, Robinson, Ronca, Sayfutyarova, Scheurer, Schurkus, Smith, Sun, Sun,
  Upadhyay, Wagner, Wang, White, Whitfield, Williamson, Wouters, Yang, Yu, Zhu,
  Berkelbach, Sharma, Sokolov, and Chan]{sun2020}
Sun,~Q. \latin{et~al.}  Recent developments in the {PySCF} program package.
  \emph{The Journal of Chemical Physics} \textbf{2020}, \emph{153},
  024109\relax
\mciteBstWouldAddEndPuncttrue
\mciteSetBstMidEndSepPunct{\mcitedefaultmidpunct}
{\mcitedefaultendpunct}{\mcitedefaultseppunct}\relax
\EndOfBibitem
\bibitem[Sun(2015)]{sun2015}
Sun,~Q. Libcint: {An} efficient general integral library for {Gaussian} basis
  functions. \emph{Journal of Computational Chemistry} \textbf{2015},
  \emph{36}, 1664--1671, \_eprint:
  https://onlinelibrary.wiley.com/doi/pdf/10.1002/jcc.23981\relax
\mciteBstWouldAddEndPuncttrue
\mciteSetBstMidEndSepPunct{\mcitedefaultmidpunct}
{\mcitedefaultendpunct}{\mcitedefaultseppunct}\relax
\EndOfBibitem
\bibitem[Schirmer \latin{et~al.}(1993)Schirmer, Trofimov, Randall, Feldhaus,
  Bradshaw, Ma, Chen, and Sette]{schirmer1993}
Schirmer,~J.; Trofimov,~A.~B.; Randall,~K.~J.; Feldhaus,~J.; Bradshaw,~A.~M.;
  Ma,~Y.; Chen,~C.~T.; Sette,~F. \textit{{K}} -shell excitation of the water,
  ammonia, and methane molecules using high-resolution photoabsorption
  spectroscopy. \emph{Physical Review A} \textbf{1993}, \emph{47},
  1136--1147\relax
\mciteBstWouldAddEndPuncttrue
\mciteSetBstMidEndSepPunct{\mcitedefaultmidpunct}
{\mcitedefaultendpunct}{\mcitedefaultseppunct}\relax
\EndOfBibitem
\bibitem[Brabec \latin{et~al.}(2012)Brabec, Bhaskaran-Nair, Govind, Pittner,
  and Kowalski]{brabec2012}
Brabec,~J.; Bhaskaran-Nair,~K.; Govind,~N.; Pittner,~J.; Kowalski,~K.
  Communication: {Application} of state-specific multireference coupled cluster
  methods to core-level excitations. \emph{The Journal of Chemical Physics}
  \textbf{2012}, \emph{137}, 171101\relax
\mciteBstWouldAddEndPuncttrue
\mciteSetBstMidEndSepPunct{\mcitedefaultmidpunct}
{\mcitedefaultendpunct}{\mcitedefaultseppunct}\relax
\EndOfBibitem
\bibitem[Sen \latin{et~al.}(2013)Sen, Shee, and Mukherjee]{sen2013}
Sen,~S.; Shee,~A.; Mukherjee,~D. A study of the ionisation and excitation
  energies of core electrons using a unitary group adapted state universal
  approach. \emph{Molecular Physics} \textbf{2013}, \emph{111}, 2625--2639,
  Publisher: Taylor \& Francis \_eprint:
  https://doi.org/10.1080/00268976.2013.802384\relax
\mciteBstWouldAddEndPuncttrue
\mciteSetBstMidEndSepPunct{\mcitedefaultmidpunct}
{\mcitedefaultendpunct}{\mcitedefaultseppunct}\relax
\EndOfBibitem
\bibitem[Dutta \latin{et~al.}(2014)Dutta, Gupta, Vaval, and Pal]{dutta2014}
Dutta,~A.~K.; Gupta,~J.; Vaval,~N.; Pal,~S. Intermediate {Hamiltonian} {Fock}
  {Space} {Multireference} {Coupled} {Cluster} {Approach} to {Core}
  {Excitation} {Spectra}. \emph{Journal of Chemical Theory and Computation}
  \textbf{2014}, \emph{10}, 3656--3668, Publisher: American Chemical
  Society\relax
\mciteBstWouldAddEndPuncttrue
\mciteSetBstMidEndSepPunct{\mcitedefaultmidpunct}
{\mcitedefaultendpunct}{\mcitedefaultseppunct}\relax
\EndOfBibitem
\bibitem[Myhre \latin{et~al.}(2018)Myhre, Wolf, Cheng, Nandi, Coriani, Gühr,
  and Koch]{myhre2018a}
Myhre,~R.~H.; Wolf,~T. J.~A.; Cheng,~L.; Nandi,~S.; Coriani,~S.; Gühr,~M.;
  Koch,~H. A theoretical and experimental benchmark study of core-excited
  states in nitrogen. \emph{The Journal of Chemical Physics} \textbf{2018},
  \emph{148}, 064106\relax
\mciteBstWouldAddEndPuncttrue
\mciteSetBstMidEndSepPunct{\mcitedefaultmidpunct}
{\mcitedefaultendpunct}{\mcitedefaultseppunct}\relax
\EndOfBibitem
\bibitem[Domke \latin{et~al.}(1990)Domke, Xue, Puschmann, Mandel, Hudson,
  Shirley, and Kaindl]{domke1990}
Domke,~M.; Xue,~C.; Puschmann,~A.; Mandel,~T.; Hudson,~E.; Shirley,~D.~A.;
  Kaindl,~G. Carbon and oxygen {K}-edge photoionization of the {CO} molecule.
  \emph{Chemical Physics Letters} \textbf{1990}, \emph{173}, 122--128\relax
\mciteBstWouldAddEndPuncttrue
\mciteSetBstMidEndSepPunct{\mcitedefaultmidpunct}
{\mcitedefaultendpunct}{\mcitedefaultseppunct}\relax
\EndOfBibitem
\bibitem[Ma \latin{et~al.}(1991)Ma, Chen, Meigs, Randall, and Sette]{ma1991}
Ma,~Y.; Chen,~C.~T.; Meigs,~G.; Randall,~K.; Sette,~F. High-resolution
  {K}-shell photoabsorption measurements of simple molecules. \emph{Physical
  Review A} \textbf{1991}, \emph{44}, 1848--1858, Publisher: American Physical
  Society\relax
\mciteBstWouldAddEndPuncttrue
\mciteSetBstMidEndSepPunct{\mcitedefaultmidpunct}
{\mcitedefaultendpunct}{\mcitedefaultseppunct}\relax
\EndOfBibitem
\bibitem[Carbone \latin{et~al.}(2019)Carbone, Cheng, Myhre, Matthews, Koch, and
  Coriani]{carbone2019}
Carbone,~J.~P.; Cheng,~L.; Myhre,~R.~H.; Matthews,~D.; Koch,~H.; Coriani,~S.
  \emph{An analysis of the performance of coupled cluster methods for core
  excitations and core ionizations using standard basis sets}; 2019; Vol.~79;
  pp 241--261, arXiv:1908.03635 [physics]\relax
\mciteBstWouldAddEndPuncttrue
\mciteSetBstMidEndSepPunct{\mcitedefaultmidpunct}
{\mcitedefaultendpunct}{\mcitedefaultseppunct}\relax
\EndOfBibitem
\bibitem[Guther \latin{et~al.}(2018)Guther, Dobrautz, Gunnarsson, and
  Alavi]{Guther2018}
Guther,~K.; Dobrautz,~W.; Gunnarsson,~O.; Alavi,~A. Time {Propagation} and
  {Spectroscopy} of {Fermionic} {Systems} {Using} a {Stochastic} {Technique}.
  \emph{Physical Review Letters} \textbf{2018}, \emph{121}\relax
\mciteBstWouldAddEndPuncttrue
\mciteSetBstMidEndSepPunct{\mcitedefaultmidpunct}
{\mcitedefaultendpunct}{\mcitedefaultseppunct}\relax
\EndOfBibitem
\bibitem[Haegeman \latin{et~al.}(2011)Haegeman, Cirac, Osborne, Pižorn,
  Verschelde, and Verstraete]{haegeman2011}
Haegeman,~J.; Cirac,~J.~I.; Osborne,~T.~J.; Pižorn,~I.; Verschelde,~H.;
  Verstraete,~F. Time-{Dependent} {Variational} {Principle} for {Quantum}
  {Lattices}. \emph{Physical Review Letters} \textbf{2011}, \emph{107}, 070601,
  Publisher: American Physical Society\relax
\mciteBstWouldAddEndPuncttrue
\mciteSetBstMidEndSepPunct{\mcitedefaultmidpunct}
{\mcitedefaultendpunct}{\mcitedefaultseppunct}\relax
\EndOfBibitem
\bibitem[Haegeman \latin{et~al.}(2016)Haegeman, Lubich, Oseledets,
  Vandereycken, and Verstraete]{haegeman2016}
Haegeman,~J.; Lubich,~C.; Oseledets,~I.; Vandereycken,~B.; Verstraete,~F.
  Unifying time evolution and optimization with matrix product states.
  \emph{Physical Review B} \textbf{2016}, \emph{94}, 165116, Publisher:
  American Physical Society\relax
\mciteBstWouldAddEndPuncttrue
\mciteSetBstMidEndSepPunct{\mcitedefaultmidpunct}
{\mcitedefaultendpunct}{\mcitedefaultseppunct}\relax
\EndOfBibitem
\bibitem[Xu \latin{et~al.}(2022)Xu, Xie, Xie, Schollwöck, and Ma]{xu2022}
Xu,~Y.; Xie,~Z.; Xie,~X.; Schollwöck,~U.; Ma,~H. Stochastic {Adaptive}
  {Single}-{Site} {Time}-{Dependent} {Variational} {Principle}. \emph{JACS Au}
  \textbf{2022}, \emph{2}, 335--340, Publisher: American Chemical Society\relax
\mciteBstWouldAddEndPuncttrue
\mciteSetBstMidEndSepPunct{\mcitedefaultmidpunct}
{\mcitedefaultendpunct}{\mcitedefaultseppunct}\relax
\EndOfBibitem
\bibitem[Yang and White(2020)Yang, and White]{yang2020}
Yang,~M.; White,~S.~R. Time-dependent variational principle with ancillary
  {Krylov} subspace. \emph{Physical Review B} \textbf{2020}, \emph{102},
  094315, Publisher: American Physical Society\relax
\mciteBstWouldAddEndPuncttrue
\mciteSetBstMidEndSepPunct{\mcitedefaultmidpunct}
{\mcitedefaultendpunct}{\mcitedefaultseppunct}\relax
\EndOfBibitem
\bibitem[Ronca \latin{et~al.}(2017)Ronca, Li, Jimenez-Hoyos, and
  Chan]{ronca2017}
Ronca,~E.; Li,~Z.; Jimenez-Hoyos,~C.~A.; Chan,~G. K.-L. Time-step targeting
  time-dependent and dynamical density matrix renormalization group algorithms
  with ab initio {Hamiltonians}. \emph{Journal of Chemical Theory and
  Computation} \textbf{2017}, \emph{13}, 5560--5571, arXiv:1706.09537
  [cond-mat, physics:physics]\relax
\mciteBstWouldAddEndPuncttrue
\mciteSetBstMidEndSepPunct{\mcitedefaultmidpunct}
{\mcitedefaultendpunct}{\mcitedefaultseppunct}\relax
\EndOfBibitem
\bibitem[Grundner \latin{et~al.}(2024)Grundner, Westhoff, Kugler, Parcollet,
  and Schollwöck]{grundner2024}
Grundner,~M.; Westhoff,~P.; Kugler,~F.~B.; Parcollet,~O.; Schollwöck,~U.
  Complex time evolution in tensor networks and time-dependent {Green}'s
  functions. \emph{Physical Review B} \textbf{2024}, \emph{109}, 155124,
  Publisher: American Physical Society\relax
\mciteBstWouldAddEndPuncttrue
\mciteSetBstMidEndSepPunct{\mcitedefaultmidpunct}
{\mcitedefaultendpunct}{\mcitedefaultseppunct}\relax
\EndOfBibitem
\end{mcitethebibliography}

\end{document}